\documentclass{aa}  

\usepackage{graphicx}
\usepackage{hyperref}
\usepackage{txfonts}
\usepackage{lipsum}
\usepackage{subcaption}   
\usepackage{ulem}
\usepackage{xcolor}
\usepackage{lscape}             
\usepackage{placeins}           
                                
\usepackage{multirow}
\renewcommand{\linenumbers}{}
\begin{document}

   \title{The Sun as an X-ray star V.}
   \subtitle{A new method to retrieve coronal filling factors}

\newcommand{\sal}[1]{\textcolor{magenta}{ #1}}



   \author{W. M. Joseph\inst{1}
        \and B. Stelzer\inst{1}
        \and S. Orlando\inst{2}
        \and M. Klawin\inst{1}
        }

   \institute{Institut für Astronomie \& Astrophysik, Eberhard Karls Universität Tübingen, Sand 1, 72076 Tübingen, Germany\\
             \email{joseph@astro.uni-tuebingen.de}
            \and INAF - Osservatorio Astronomico di Palermo, Piazza del Parlamento 1, 90134 Palermo, Italy\\ }

   \date{Received September 30, 20XX}

 
  \abstract
    {Stellar coronae are unresolved in X-rays, so inferences about their structure rely on spectral analysis. The “Sun-as-an-X-ray-star’’ (SaXS) approach uses the Sun as a spatially-resolved template to interpret stellar spectra, but previous SaXS implementations were indirect and computationally heavy.}
    {We present a new SaXS implementation that converts solar emission measure distributions (EMDs) of distinct coronal region types into {\sc XSPEC} spectral components, and we test whether broad-band X-ray spectra alone can recover the filling factors of those region types.}
    {We built {\sc XSPEC} multi-temperature spectral models for four solar region types (background/quiet corona, active regions, cores and flares by considering EMDs derived from the analysis of {\it Yohkoh}/SXT data and by translating each EMD bin to an isothermal {\sc apec} component. These {\sc XSPEC} models were fit (using PyXspec) to two one-hour DAXSS spectra representative of quiescent (2022-06-29) and flaring (2022-04-25) states. Best-fit normalizations were converted into projected areas and filling factors and compared with near-coincident {\it Hinode}/XRT full-disk images for validation.}
    {Using the {\it Yohkoh}/SXT EMDs, we found that the spectrum of the Quiescent Sun is dominated by active region emission (filling factor $\approx 22$\%), with the background corona poorly constrained, while the spectrum of the Flaring Sun is best described by a combination of active regions, cores, and flares with filling factors $\approx 47.5$\%, $\approx 4.1$\%, $\approx 0.062$\%, respectively. We checked  that the dominant components qualitatively match spatial features in {\it Hinode}/XRT images. Major limitations are the DAXSS low-energy calibration cutoff ($\sim 0.7$ keV) and the small, non-uniform {\it Yohkoh} EMD sample adopted, which may affect constraints on cool, low-emission regions and on elemental line emission.}
    {We demonstrated that our SaXS implementation enables direct retrieval of coronal filling factors from broad-band X-ray spectra and provides a physically motivated alternative to ad-hoc few-temperature fits. This approach can therefore be routinely applied to stellar X-ray spectra to infer the distribution of coronal structures.}


   \maketitle


\section{Introduction}
\label{sect:intro}

High-energy observations are crucial for understanding solar and stellar coronae, where plasma is magnetically confined and heated to millions of kelvin \citep{Pallavacini1981}. The solar corona hosts a variety of magnetic structures that appear bright and hot in X-rays \citep{Reale2014}. These structures play a central role in shaping solar activity and in addressing the long-standing coronal heating problem \citep{Klimchuk2006,Klimchuk2015}. In contrast, for stars other than the Sun, X-ray emission remains spatially unresolved: stellar coronae appear as point sources whose spectra represent the integrated contribution of regions spanning a wide range of temperatures and densities. Interpreting such spectra therefore relies on indirect approaches that link the observed emission to the underlying physical structures on the stellar surface.


The Sun provides the unique opportunity to bridge this gap. As the only star whose corona can be spatially resolved, it can serve as a reference for stellar coronal studies through ``Sun-as-a-star'' methods that attempt to answer: How would the Sun appear observationaly if it were a distant star? (e.g., \citealt{Peres2000, Livingston2007, Cameron2019}). Using detailed solar observations enabled by the proximity of the Sun, we can infer the ongoings on the surface and atmosphere of distant stars of similar type, that is late-type magnetically active stars. 

The Sun-as-an-X-ray-Star (SaXS) method \citep{Orlando2000, Peres2000, Reale2001} is the application of this concept to stellar coronae as observed in the X-ray band. It was motivated by the observation that the Sun consists of various types of magnetically active regions whose brightness, temperature, and density increase in the following order:  
(i) Background corona (BKC), representing the quiet corona,  
(ii) Active regions (AR), consisting of magnetic loops that trap plasma at temperatures of a few MK, 
(iii) Cores of active regions (CO), composed of hot, dense magnetic loops, and  
(iv) Flares (FL), explosive energy releases occurring in coronal loops resulting from magnetic reconnection. These coronal region types were originally demarcated only by their pixel intensities, in units of DN/s, on images obtained with the Soft X-ray Telescope (SXT; \citealt{Tsuneta1991}) onboard the \textit{Yohkoh} satellite \citep{Ogawara1991}. The abundance of the different types of these coronal regions is a marker of solar activity, and their variations are used to study the solar cycle (e.g., \citealt{Peres2000, Orlando2001}). 

The different types of magnetically active structures on the Sun have distinct X-ray emission measure distributions (EMDs).  These distributions,  as observed by {\it Yohkoh}/SXT,  were presented by \citet{Orlando2001, Orlando2004} for the BKC, AR, and CO regions, and by \citet{Reale2001} for individual flares of the following GOES classes: C5.8, M1.0, M1.1, M2.8, M4.2, M7.6, X1.5, and X9.0.

In the  first applications of the SaXS method, these {\it Yohkoh} EMDs were used to predict the occurrence and amount of the different solar coronal region types on HD\,81809 \citep{Favata2008, Orlando2017}, $\epsilon$\,Eri \citep{Coffaro2020} and Kepler\,63 \citep{Coffaro2022}. This was achieved by an approach that involved creating synthetic total emission measures, $EM_{\rm tot}$,  by scaling the EMDs of the individual  coronal region types,  $EMD_{\rm reg}$, by their filling factors, $ff_{\rm reg}$, and summing them,  $EM_{\rm tot} = \sum_{\rm reg}{ff_{\rm reg} EMD_{\rm reg}}$. The filling factors represent the percentage of the stellar corona covered by different types of magnetic region.
For each total emission measure distribution, defined by the filling factors of the various magnetic region types, an X-ray spectrum was synthesized. These synthetic spectra were then analyzed as if they were observational data: fitted in {\sc XSPEC} with thermal models to derive the coronal properties, namely, the X-ray temperature ($T_{\rm x}$) and luminosity ($L_{\rm x}$). By comparing the synthetic ($T_{\rm x}$, $L_{\rm x}$) pairs with the corresponding values obtained from the same thermal model to the observed stellar X-ray spectrum, we identified the combination of filling factors that best reproduces the observation. This allowed us to infer the relative contributions and importance of the different coronal regions in the target star.

In its first application to the solar-like star HD\,81809 \cite{Favata2008} synthesized a corona composed of solar AR and CO, which \citet{Orlando2004} had shown to characterize the solar corona throughout its activity cycle.
The study found that the X-ray temperature and luminosity predicted from the AR and CO filling factors characteristic of the solar cycle bracketed the observed values ($T_{\rm x}$, $L_{\rm x}$) of HD\,81809. A subsequent study of the same star, based on an approximately doubled {\it XMM-Newton} monitoring time, revealed a wider scatter of the star in the ($T_{\rm x}$, $L_{\rm x}$) plane \citep{Orlando2017}. Consequently, \citet{Orlando2017}, along with later studies on other stars (\citealt{Coffaro2020, Coffaro2022}), computed a full grid of synthetic spectra — based on a corresponding grid of filling factors — and compared the resulting best-fit parameters with the observed stellar properties.

For $\epsilon$\,Eri this approach has revealed a high (60-90\%) surface coverage with magnetic regions throughout the whole activity cycle   \citep{Coffaro2020}. This finding was considered as a potential origin for the saturation phenomenon \citep{Fleming1989}. In this interpretation, increasing the stellar spin rate beyond a certain threshold does not increase the star's X-ray  luminosity because the corona is already fully covered with magnetic structures and can not accommodate additional emitting regions.

These earlier  applications of the SaXS method to stellar X-ray spectra, thus, demonstrated its high scientific value for quantifying the solar-stellar connection paradigm. However, although the method provided  reasonable estimates of the filling factors and far-reaching results  on the structure of stellar coronae, the previous implementation of the SaXS method, it is  complex, indirect, and computationally demanding.

We present here a new, more efficient implementation of the SaXS technique. It consists in converting the EMDs of solar coronal regions derived from \textit{Yohkoh} images into empirical X-ray spectral models that can be directly used to fit stellar X-ray spectra. The aim to retrieve the filling factors of the corona of a given star with each of these solar coronal region types is then achieved by weighting the contribution of the model of each coronal region type in the fitting and finding the best  combination of them.

However, before applying this technique to distant unresolved stars it needs to be calibrated on solar spectra. This requires solar X-ray spectra covering the continuous range of energies (0.2–10 keV) observed with astrophysical facilities like the X-ray Multi-Mirror Mission \citep[{\it XMM-Newton};][]{Jansen2001} or the extended ROentgen Survey with an Imaging Telescope Array \citep[eROSITA;][]{Predehl2021} on board the Russian Spektrum-Roentgen-Gamma mission \citep[SRG;][]{Sunyaev2021}, which perform X-ray spectroscopy of other stars. Such solar broad-band soft X-ray spectra have been lacking until recently.

Beginning in 2016, the Miniature X-ray Solar Spectrometer CubeSat-1 (MinXSS-1; \citep{Mason2016}), its successor MinXSS-2 \citep{Mason2020}, and a sounding rocket flight of the Dual-zone Aperture X-ray Solar Spectrometer (DAXSS; \citep{Woods2017,Woods2023}) provided the first solar spectra in the 0.5–10 keV band. More recently, DAXSS is being operated on board the INSPIRESat-1 satellite \citep{Chandran2021}, delivering continuous observations suitable for testing our method.

In this paper, we present a new implementation of the SaXS method and apply it to solar X-ray spectra from DAXSS/INSPIRESat-1, covering two different levels of solar activity. We verify the consistency between spectrally derived coronal regions and their filling factors using near-simultaneous full-disk {\it Hinode}/XRT images. In Sect.~\ref{sec:model-creation}, we describe the construction of {\sc XSPEC} spectral models from EMDs of different coronal region types and the retrieval of filling factors from spectral fits. In Sect.~\ref{sec:solar-data}, we present their application to DAXSS data. In Sect.~\ref{sec:discussion}, we discuss the results, and finally, in Sect.~\ref{sec:conclusions}, we provide our conclusions.

\section{Definition of {\sc XSPEC} models for the SaXS method}
\label{sec:model-creation}

In this section, we describe how we obtain coronal region spectral models from the solar emission measure distributions observed with {\it Yohkoh}/SXT.
Specifically, we convert the emission measure distribution of each solar coronal region $EM(T)_{reg}$ observed with {\it Yohkoh} into a corresponding empirical spectral model that we then implement in PyXSPEC \citep{Gordon2021}.

\begin{figure} 
\centering
\includegraphics[width=\columnwidth]{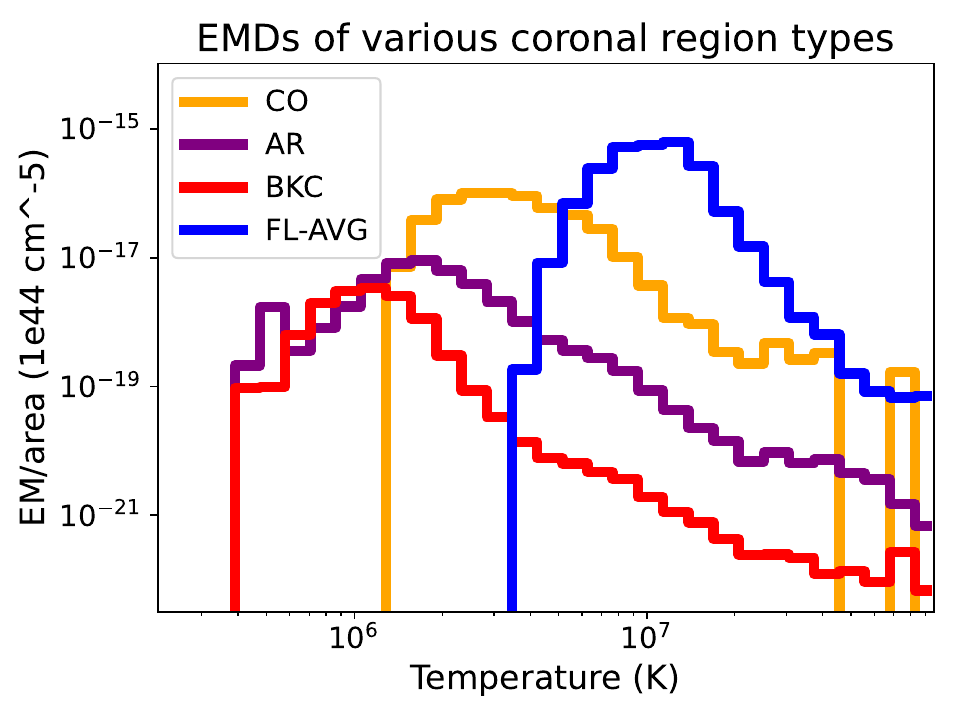}
\caption{EMDs of the BKC, AR, CO and FL-AVG regions as described in Sect.~\ref{sect:intro}. The y-axis shows the emission measure per unit area, 
in units of $10^{44}$ cm$^{-5}$. All EMDs shown here represent time-averaged distributions constructed from multiple snapshots of flares or other types of regions.}
\label{fig:em-dist}
\end{figure}

The EMDs per unit projected area derived from {\it Yohkoh}/SXT data by \citet{Orlando2001}, \citet{Orlando2004} and \citet{Reale2001} for the coronal structures described in Sect.~\ref{sect:intro} are shown in Fig.\ref{fig:em-dist}. These EMDs were derived from a limited set of {\it Yohkoh}/SXT observations, namely 23 synoptic images containing various region types \citep{Orlando2001}, 78 observations following a single active region through its evolution \citep{Orlando2004}, and 8 flares covering different phases of their evolution from rise to peak and decay \citep{Reale2001}.

\citet{Reale2001} derived EMDs for multiple snapshots of each flare. In the present work, to simplify the 
development of the new SaXS methodology in this initial phase and to avoid the complications introduced by temporal variations, we construct time-averaged flare EMDs by averaging the EMDs from all snapshots of each flare. 

\noindent

\begin{figure} 
\centering
\includegraphics[width=\columnwidth]{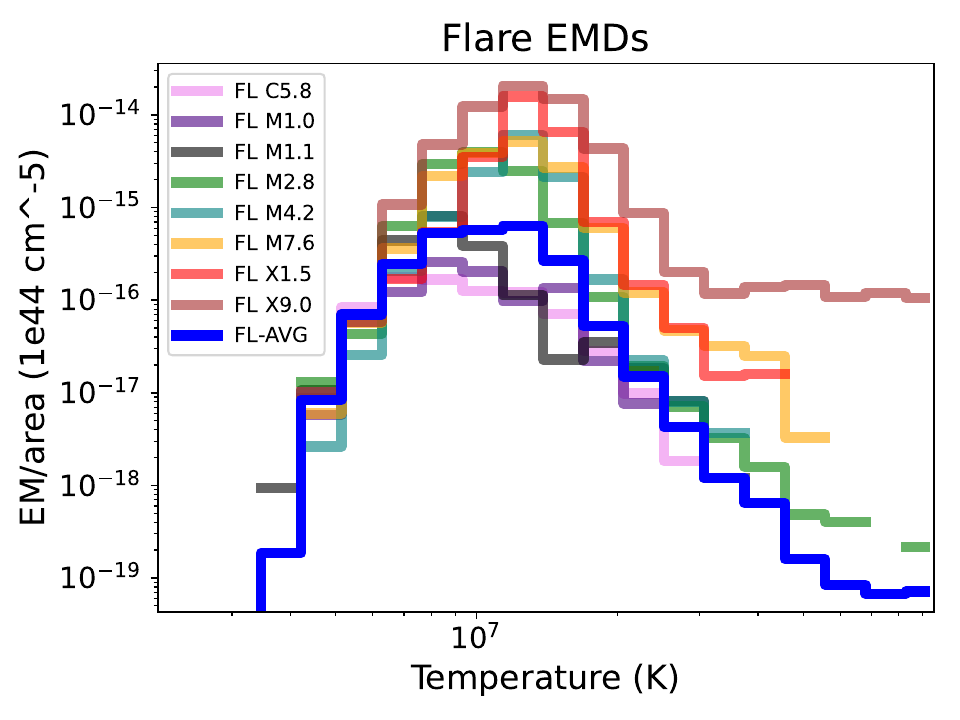}
\caption{EMDs of the C5.8, M1.0, M1.1, M2.8, M4.2, M7.6, and X1.5 flares (individual colors) along with their weighted average EMD (FL-AVG, blue). The y-axis shows the emission measure per unit area, 
in units of $10^{44}$ cm$^{-5}$.}.
\label{fig:emd-flare-avg}
\end{figure}

The EMDs of the individual flares (C5.8, M1.0, M1.1, M2.8, M4.2, M7.6, X1.5, and X9.0) are shown in Fig.~\ref{fig:emd-flare-avg}.For simplicity, we combine the EMDs of all the flares into a single representative "average flare" EMD (hereafter FL-AVG) When constructing FL-AVG, we accounted for the fact that flares with different peak amplitudes (which define their GOES class) occur at different frequencies. Accordingly, FL-AVG was obtained as a weighted average of the individual flare EMDs, where the weights were determined by the slope of the cumulative flare energy frequency distribution (FFD) in logarithmic space, $\alpha$, following $N(E) \propto E^{\alpha}$. Here, $E$ denotes the flare energy, and we adopted $\alpha = -1.54$, which is the FFD slope derived for solar X-ray flares from {\it Yohkoh} observations \citep{Aschwanden2002}. The energy of the eight individual flares was estimated using the empirical relation between the GOES peak flux, $F_{\rm GOES}$\footnote{These fluxes, by definition, determine the GOES class of the flares.}, and the corresponding white-light flare (WLF)\footnote{White-light flares are characterized by a sudden enhancement of optical continuum emission \citep{Svestka1970,Neidig1989,Song2018}.} energy, $E_{\rm WLF}$ from \citet{Namekata2017}. This relation is expressed as $\log(E_{WLF})=a+b\cdot \log(F_{GOES})$, with coefficients $a = 33.67$ and $b = 0.87$. The resulting average FL-AVG EMD is shown in blue in Figs.~\ref{fig:em-dist} and \ref{fig:emd-flare-avg}.

To derive a spectral model from each of the EMDs, we assume that each bin $i$ in the EMD of a given type of coronal region ($T_{\rm i,reg}, EM_{\rm i,reg}$) represents emission from an isothermal plasma. The total EMD of that type of region is then described by a  distinct distribution of several isothermal plasmas effectively constituting a multi-temperature plasma. Such a multi-temperature plasma model can be constructed in {\sc XSPEC} \citep{Arnaud1996} with a sum of $i$ {\sc apec} (or {\sc vapec} for variable abundances) components \citep{Smith2001}, each characterized by ($T_{\rm i,reg}$, $Norm_{\rm i,reg}$), as well as the coronal abundances as input parameters. We implement this modeling approach using PyXspec \citep{Gordon2021}, a Python package that provides an interface to {\sc XSPEC}.  The abundances are here defined for the star, i.e. they are assumed to be the same throughout the whole corona independent of the type of magnetic structure. Therefore, each coronal region model $M_{\rm reg}$ is defined as
\begin{equation}
\label{eq:mreg}
M_{\rm reg}=\sum_{i}{APEC(T_{\rm i,reg}, Norm_{\rm i,reg}, Abund_{\rm star})}
\end{equation}
\noindent
where $M_{\rm reg}$ can be $M_{\rm BKC}$, $M_{\rm AR}$, $M_{\rm CO}$ or $M_{\rm FL-AVG}$. The spectral models for all four region types and their individual {\sc apec} components are visualized in Appendix~\ref{fig:model-apec-comps}.

The normalizations, $Norm_{i,reg}$, are fixed values obtained from the $EM_{i,reg}$ values of the region type's EMD using the relation
\begin{equation}
\label{eq:norm}
    Norm_{i,reg}=10^{-14} \times \frac{EM_{i,reg}}{4 \pi D^{2}}
\end{equation}

\noindent
where D is the distance to the star in cm.
The elemental abundances can either be set as a global value, expressed in units of solar abundances using the {\sc apec} model, or customized for each chemical element individually using the {\sc apec} variants {\sc vapec} and {\sc vvapec}, which are also available in {\sc XSPEC}. In principle, {\sc vapec} models would be more appropriate when the elemental abundances are already known for different temperatures on the Sun, or when applying the models to stars whose coronal abundances are expected to differ from the solar values. However, since our primary goal is to determine filling factors rather than to derive elemental abundances, we adopt the standard {\sc apec} model with fixed quiescent solar coronal abundances. Using a model with individual elemental  abundances would vastly increase the number of free parameters to be fit, which detracts from our primary goal of finding the filling factor, for which we only need the normalization factor of the spectral fit, as will be described in the following paragraphs.

Solar and stellar broad-band soft X-ray spectra can then be fit with a spectral model, $M_{\rm spec},$ that is an additive combination of the spectral models for various types of regions. Hereby, the contribution from a given region, $M_{\rm reg}$ as a whole can be scaled up or down during spectral fitting, with the scaling measured as the total normalization of the coronal region $Norm_{\rm reg}$. The spectral model to be fitted is, thus,
\begin{equation}
\label{eq:mspec}
M_{\rm spec}=\sum_{reg}{(Norm_{\rm reg} \cdot M_{\rm reg}).}
\end{equation}

\noindent
The total emission measure of a particular type of region on the star,  $EM_{reg,tot}$,  is then just the total emission measure per unit area of that type of region on the Sun, $EM_{\rm reg,\odot}$,
\begin{equation}
\label{eq:EM-reg}
EM_{\rm reg,\odot}=\sum_{i}{EM_{i,reg}} 
\end{equation}

\noindent
scaled by $Norm_{reg}$ obtained from the spectral fit:
\begin{equation}
\label{eq:EM-reg-tot}
    EM_{reg,tot}=EM_{\rm reg,\odot}\cdot Norm_{reg}
\end{equation}

On the Sun, $Norm_{reg}$ is then simply the projected area of that coronal region in $cm^{2}$. 
Therefore, the filling factor of the region on the Sun, $ff_{reg}$, would be

\begin{equation}
\label{eq:ff-reg}
    ff_{reg}=\frac{Norm_{reg}}{Area_{\rm proj,\odot}}
\end{equation}
where $A_{\rm proj,\odot}$ is the projected area of the solar disk.  
When observing a star in X-rays, the detected emission originates not only from the visible photospheric disk but also from the extended corona above the limb. This is also true for the broad-band solar X-ray spectra from DAXSS. To account for this, we include the vertical extension of the corona in the total projected area, $A_{\rm proj,\odot}$. We achieve this by approximating the extent of the solar corona by its scale height, defined as
\begin{equation}
S_{\rm BKC} = \frac{2 k T_{\rm BKC}}{\mu m_{\rm H} g_{\odot}},
\end{equation}
where $\mu = 0.6$ is the mean molecular weight of the fully ionized coronal plasma, and $T_{\rm BKC}$ is the emission-measure–weighted average temperature of the background corona.
The corresponding effective projected area is then given as
\begin{equation}
A_{\rm proj,\odot} \simeq \pi (R_\odot + S_{\rm BKC})^{2},
\end{equation}
where $R_\odot$ is the solar radius. This formulation includes both the photospheric disk and the coronal layer extending approximately one scale height beyond the solar limb. This amounts to a total area of $A_{\rm proj,\odot} \approx 1.7\times10^{22}$ cm$^{2}$.

\section{Application of SaXS to solar data}
\label{sec:solar-data}

The new version of the SaXS method, that consists in defining XSPEC spectral models for the different coronal region types and fitting them to observed spectra, is  introduced here for the first time.  Therefore, before proceeding to applying this approach to stellar X-ray spectra, we provide in this article the first test on the Sun itself. 
The goal is analogous to that of future applications to other stars e.g. our first application to AD\,Leo (Joseph et al., in prep.), that is, to retrieve the coronal filling factors for the different types of magnetic structures. In contrast to other stars, however, the Sun offers the possibility to verify the results on actual images, namely to check whether the filling factors derived from the spectra are consistent with the structures seen at the same time in full-disk observations of the Sun. 
To this end, we apply the new SaXS methodology to soft X-ray spectra from DAXSS, obtain the best fitting combination of coronal region models (Sect.\ref{subsec:spec-fits-results}) and derive their corresponding filling factors. Subsequently, we validate the results of the spectral fits on solar {\it Hinode}/XRT images taken close in time (Sect.~\ref{sec:ff-validation-hinode}).

\subsection{DAXSS observations and selection of representative spectra}
\label{subsec:daxss-obs}

We obtained DAXSS Level 2 data stored as mission-length NCDF files on \url{lasp.colorado.edu/home/minxss/data/}. DAXSS data are updated periodically as the mission continues to downlink new observations daily. At the time of writing, version 2.1.0 was available, covering February 28, 2022 to October 24, 2023.  This time span corresponds to the middle of the rise phase of Solar Cycle No. 25.

The DAXSS data can be unpacked to provide light curves and spectra in the 0.4-12.0\,keV band. DAXSS spectra are available in native cadence (a few minutes), one-hour, and one-day averages. We used the one-hour averaged spectra as our ultimate goal with the SaXS method is to test whether solar-type regions are found on stars by fitting spectral models to stellar spectra, which are typically averaged over hour timescales to collect sufficient photon statistics for reliable spectral fits.

\begin{figure} 
\centering
\includegraphics[width=\columnwidth]{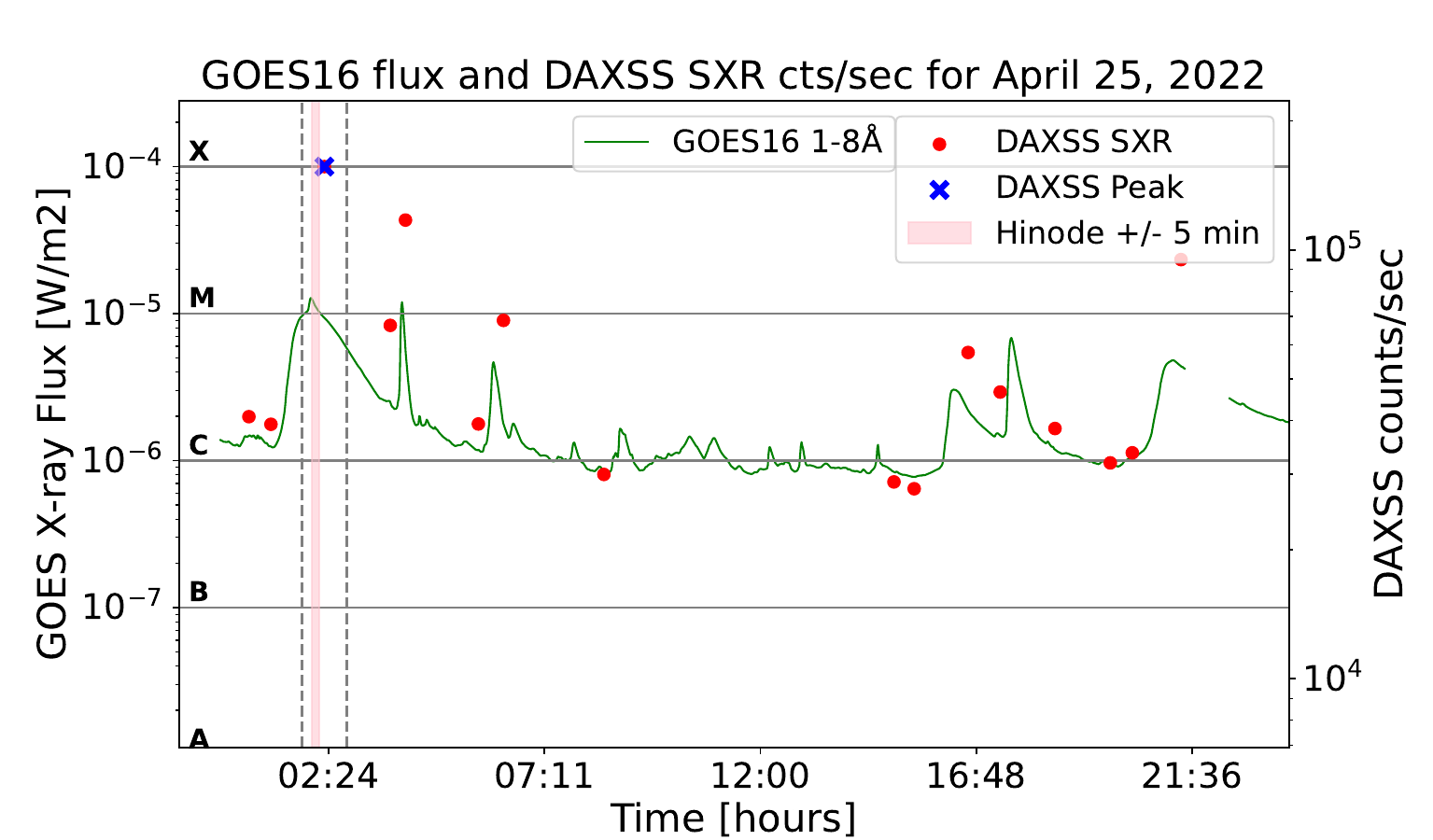}
\includegraphics[width=\columnwidth]{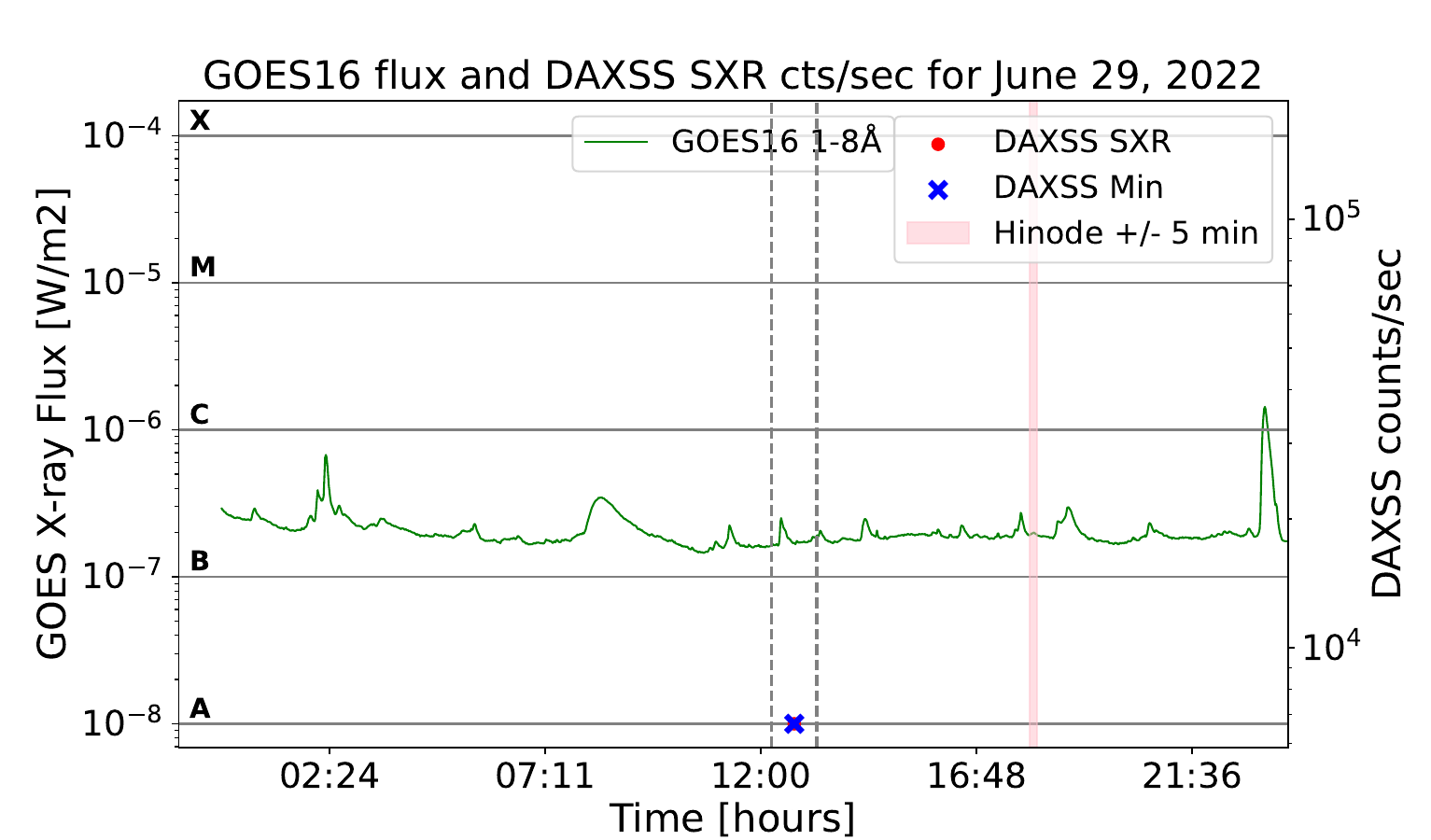}
\caption{GOES X-ray flux in the 1-8 \AA~band  (green)  and DAXSS count rates for April 25, 2022 and June 29, 2022. Horizontal grey lines demarcate the GOES flare classes A, B, C, M and X. The epochs of the DAXSS spectra selected for analysis in this paper are marked in the DAXSS light curve as blue crosses and dashed vertical lines indicate the one-hour exposure time of the spectra. The {\it Hinode} synoptic images used to verify the SaXS method correspond to the time period indicated by the pink vertical band.}
\label{fig:goes-lcs}
\end{figure}

To test the performance of the SaXS models for different solar activity states we used the DAXSS/INSPIRESat-1  mission light curve to identify the two observations with the lowest count rate and highest count rate. These correspond to June 29, 2022 at 12:44 and April 25, 2022 at 02:18 respectively, and we extracted their corresponding spectra. We refer to these hereafter as the Quiescent Sun and Flaring Sun observations, respectively. 
The GOES light curves acquired simultaneously with the DAXSS observations reveal the Sun's activity level which is clearly different for  the two observations  (see Fig.\ref{fig:goes-lcs}).  
The April 25 observation corresponds to an M class flare (in the 1-8\,\AA~band used for GOES flare classification), while the June 29 observation shows minimal activity with only a small B class flare barely rising above the surrounding quiescent background level.

\label{sec:spec-fitting}

\subsection{Refining EMDs based on spectral fitting results}
\label{subsect:refining-emd}

We applied the SaXS {\sc XSPEC} models (Sect.~\ref{sec:model-creation}) separately to the two DAXSS spectra described in Sect.~\ref{subsec:daxss-obs}. During the fitting, the elemental abundances were fixed to quiescent solar values, adopting the Feldman Standard Extended Coronal abundances (feld in {\sc XSPEC}; \citealt{Feldman1992}), with the {\sc XSPEC} parameter for the abundance set to unity.
Upper limits were imposed corresponding to a total filling factor of 100\% to ensure that the derived projected areas did not exceed the solar corona’s total projected area  defined in Sect.~\ref{sec:model-creation}. The primary quantities obtained from each fit are the filling factors, $ff_{\rm reg}$ (Eq.~\ref{eq:ff-reg}), for all region types included in the best-fitting model. 

\begin{figure} 
\centering
\includegraphics[width=\linewidth]{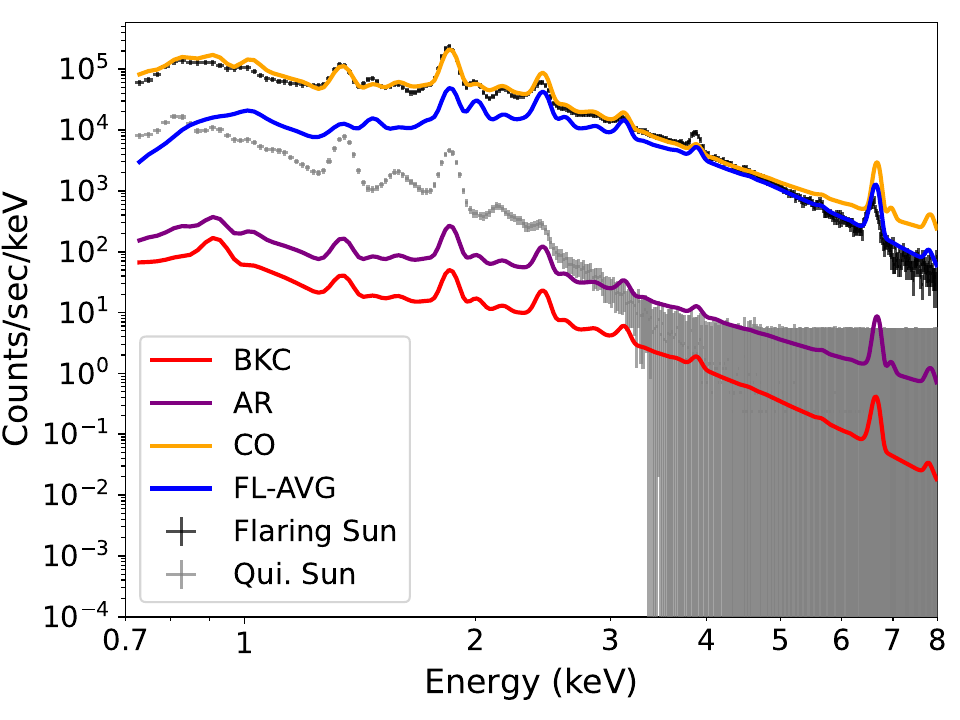}
\caption{DAXSS spectra corresponding to the lowest and highest count rates in the DAXSS light curve along with arbitrarily scaled SaXS models of the various types of coronal regions. This  demonstrates the steeper slope of the solar spectra compared with the spectrum of all types of coronal region which indicates an over-contribution of low-signal EM bins (see text in Sect.~\ref{subsect:refining-emd}}.
\label{fig:demo-all-em-bins-bad-fit}
\end{figure}

Initial fits using spectral models derived from the full EMDs reveal systematic discrepancies with the observed spectral shape. Specifically, the observed spectra exhibit steeper high-energy drop-offs than predicted by any of the SaXS models in the 0.7–3.0~keV range. This is illustrated in Fig.~\ref{fig:demo-all-em-bins-bad-fit}, in which we overlay (without fitting) the different SaXS spectral models to the two observed DAXSS spectra.
We attribute these  discrepancies to the bins at the EMD tails, in particular the high-temperature tails of CO, AR and BKC, which have low values of emission measure but seem to over-contribute to the spectral models.
We, therefore, re-define the EMD for each type of region, restricting it to bins within one order of magnitude of the peak of the EMD, as shown in Appendix~\ref{fig:em-dist-reduced}. The need for such an ad hoc adjustment indicates that the EMDs used here are not fully representative for the region types. This is most likely related to the time-evolution of the magnetic structures and their associated EMDs; see also discussion in Sect.~\ref{subsect:discussion_3}. 

We regenerate the corresponding SaXS spectral  models from the restricted EMDs. In Appendix~\ref{fig:spec-models-em-dist-reduced} we show how the spectral models for the different types of coronal region change as a result of the modification of the EMD. Clearly, the EMD bins that have now been removed (translucent bins in Appendix.\ref{fig:em-dist-reduced}) drastically increase the flux in the high-energy portion of the spectra, making them inconsistent with the observed DAXSS spectra.

\subsection{Spectral fits and results}
\label{subsec:spec-fits-results}

Using the restricted EMDs we  systematically test the individual components on the two DAXSS spectra introduced in Sect.~\ref{subsec:daxss-obs} as explained in the following.

\subsubsection{Quiescent Sun spectrum}
\label{subsubsec:spec-fitting-qui}

\begin{figure} 
\centering
\includegraphics[width=\columnwidth]{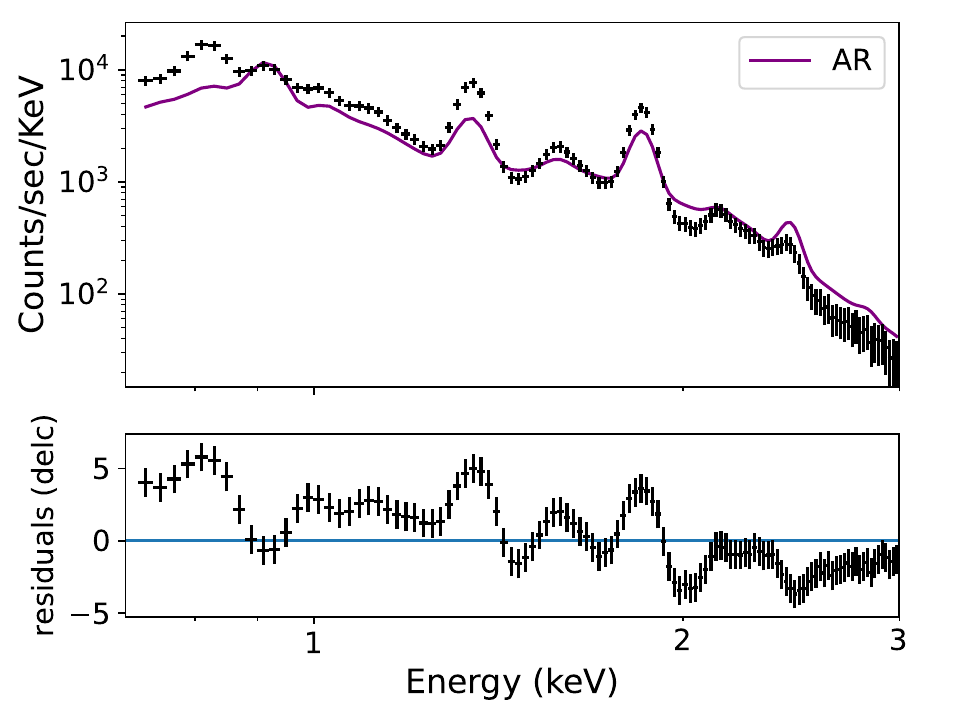}
\includegraphics[width=\columnwidth]{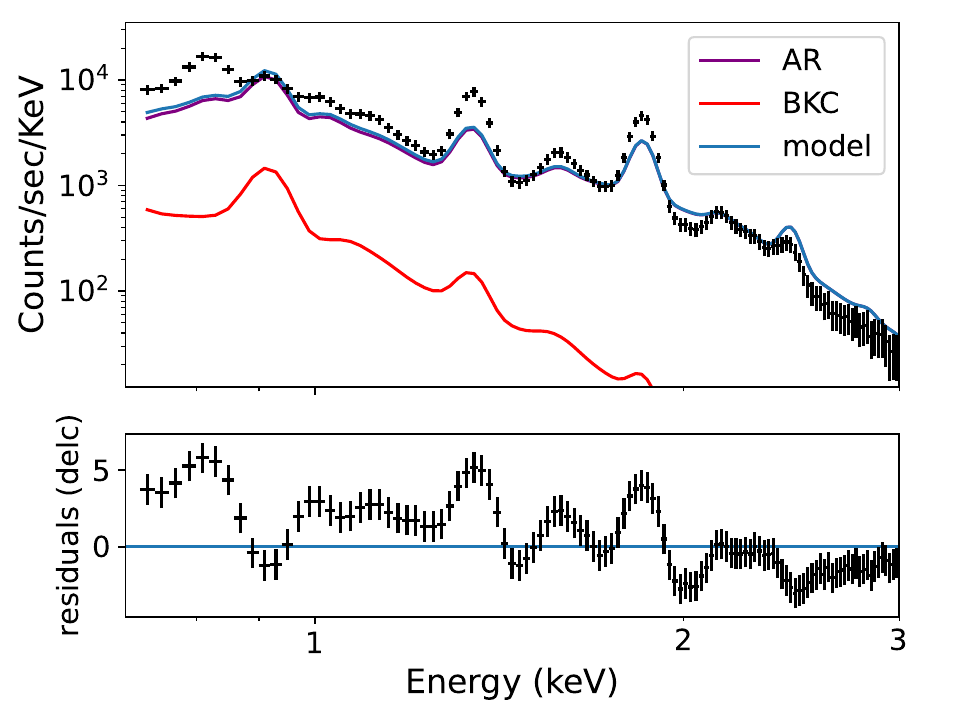}
\caption{DAXSS spectrum from 29 June 2022 representing the Quiescent Sun fitted with AR model alone (top) and with added BKC component (bottom).}
\label{fig:bkc-ar-dist-reduced-spectra}
\end{figure}

Starting by considering a single type of region, the AR model provides the best fit to the Quiescent Sun spectrum with a filling factor of $23.54\pm0.31\%$ (Fig.~\ref{fig:bkc-ar-dist-reduced-spectra}, top panel). However, low-energy residuals suggest the need for an additional lower-temperature
component. Adding the BKC model (Fig.~\ref{fig:bkc-ar-dist-reduced-spectra}, bottom panel) with the same EMD restrictions improves the spectral fit ($\chi_{red}^{2}$ = 5.33 vs. 5.78  previously). The BKC  component contributes minimally to the overall spectrum, resulting in its filling factor being poorly constrained, and the AR component filling factor changing only marginally. This two-region model has the filling factors $ff_{\rm BKC} = 94.12\pm65.85\%$ and  $ff_{\rm AR} = 21.83\pm0.4\%$. 

The BKC is poorly constrained as the emission from the much brighter AR dominates the spectrum, despite the latter having a smaller filling factor.
The remaining residuals may indicate that the spectral SaXS models require further refinement, which we leave to future work.

\subsubsection{Flaring Sun spectrum}
\label{subsubsec:spec-fitting-flare}

\begin{figure} 
\centering
\includegraphics[width=\linewidth]{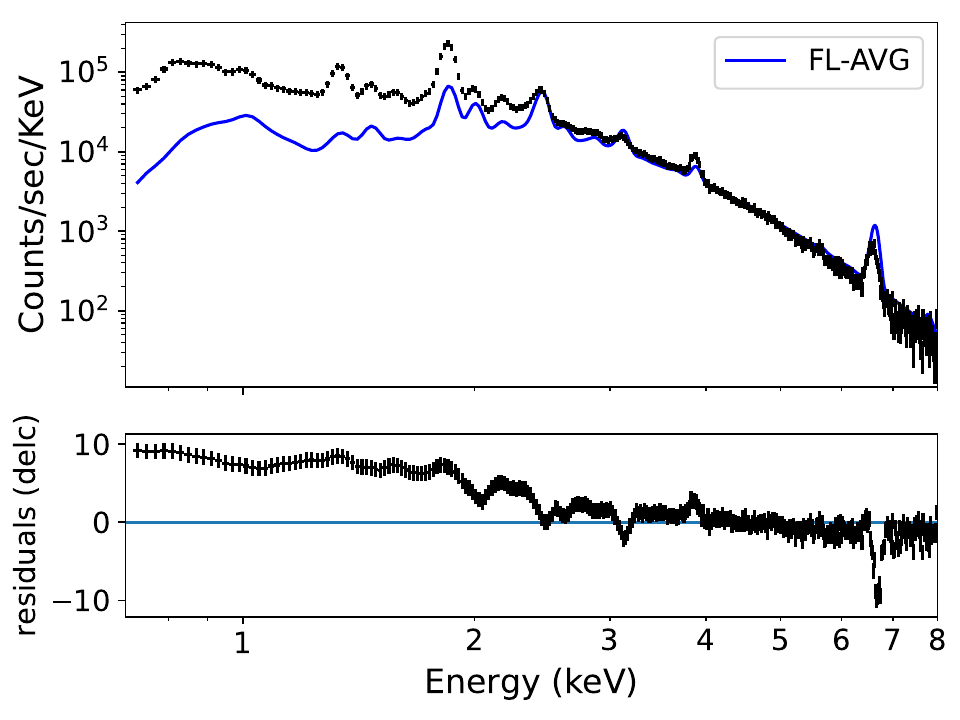}
\includegraphics[width=\linewidth]{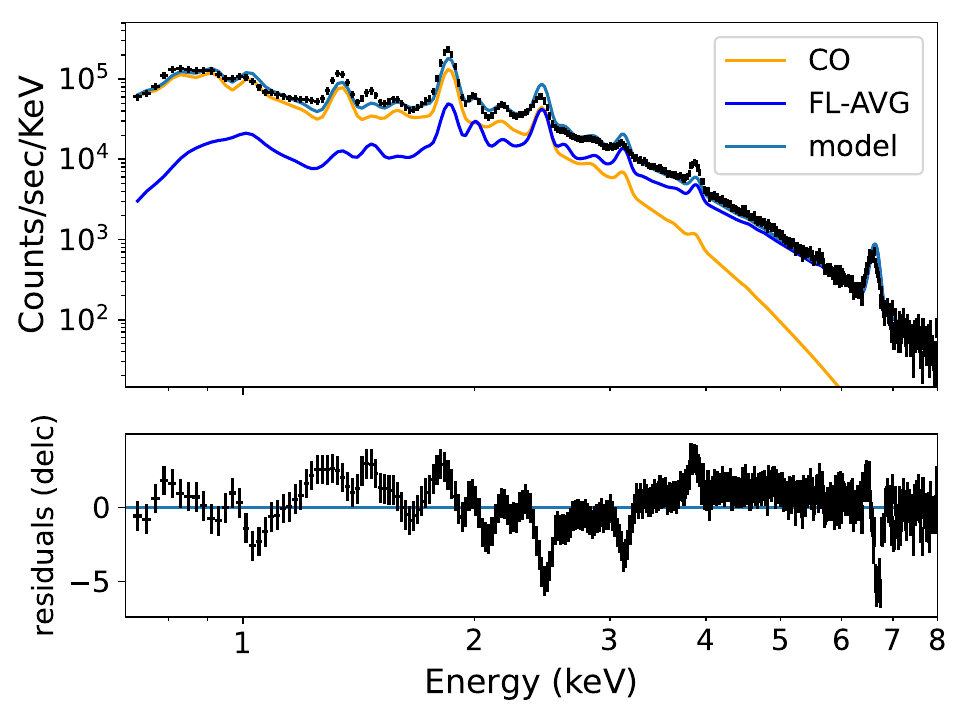}
\includegraphics[width=\linewidth]{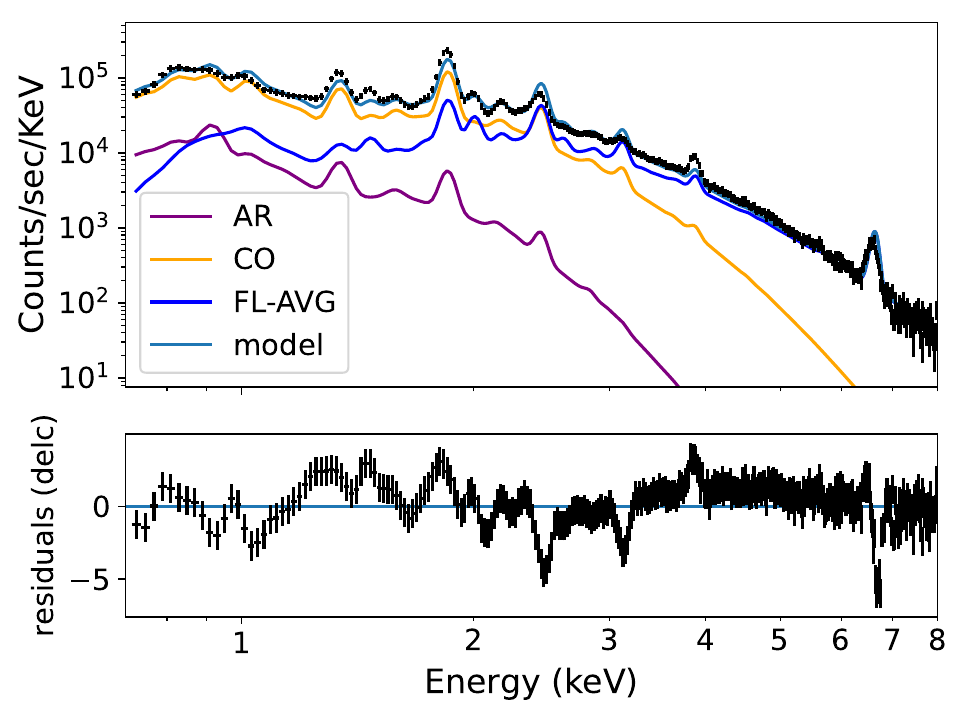}
\caption{DAXSS spectrum of 25 Apr 2022, representing a flare fitted with FL-AVG alone (top), FL-AVG + CO (middle), and FL-AVG + CO + AR (bottom).}
\label{fig:fl-obs-fit}
\end{figure}

We apply the same methodology to the flare spectrum. Given the enhanced high-energy emission compared with the Quiescent Sun spectrum (see Fig.~\ref{fig:demo-all-em-bins-bad-fit}) we start the fitting  with FL-AVG which produces  the best fit to the high energy part of the spectrum. This model fits the high-energy portion well, but significant low-energy residuals remain, yielding $\chi_{red}^{2}=13.64$  (Fig.~\ref{fig:fl-obs-fit}, top panel).

After adding CO the model successfully fits the low-energy spectrum with strongly reduced residuals, $\chi_{red}^{2}=2.29$, (Fig.~\ref{fig:fl-obs-fit} middle panel) and filling factors $ff_{\rm CO} = 4.47\pm0.069\%$ and $ff_{\rm {FL-AVG}} = 0.06\pm0.00068\%$. Since cores typically accompany active regions, we include the AR model, which provides a further marginal improvement ($\chi_{red}^{2} = 2.26$) but leaves the AR filling factor poorly constrained at $ff_{\rm AR} = 47.51\pm13.97\%$. Except for its softest part the spectrum is dominated by core and flare emission, with filling factors of $ff_{\rm CO} = 4.07\pm0.14\%$ and $ff_{\rm {FL-AVG}} = 0.062\pm0.00077\%$ in the three-component fit (Fig.~\ref{fig:fl-obs-fit}, bottom panel). Adding a BKC component does not further improve the fit ($\chi_{\mathrm{red}}^{2} = 2.27$) because the emission is dominated by the other regions with higher emission per unit area. The BKC component is very poorly constrained yielding $ff_{\mathrm{BKC}} = (5.4\pm1102)\times10^{-3}\%$.

\subsection{Validation with Hinode Observations}
\label{sec:ff-validation-hinode}

\label{subsec:hinode-obs}

\begin{figure} 
\centering
\includegraphics[width=\columnwidth]{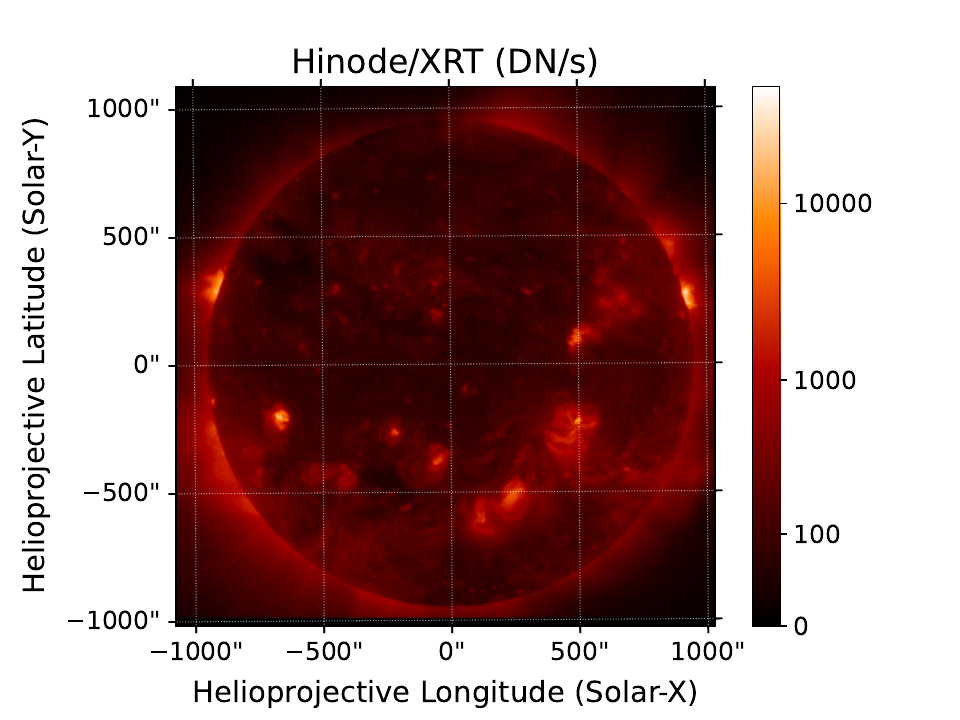}
\includegraphics[width=\columnwidth]{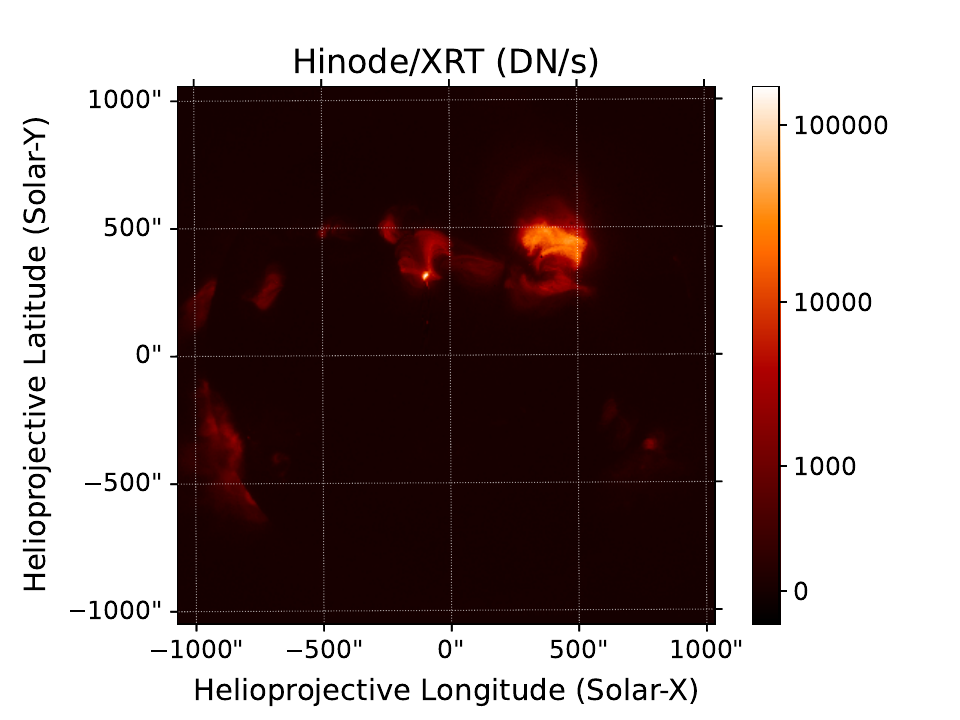}
\caption{{\it Hinode}/XRT images corresponding to Quiescent and Flaring Sun DAXSS observations (top and bottom panel respectively).}
\label{fig:hinode-full-images}
\end{figure}

To verify  if the filling factors inferred from the spectral analysis are consistent with the structures seen in contemporaneous images of the Sun we obtained the {\it Hinode}  Level 2 synoptic composite images (\url{http://solar.physics.montana.edu/HINODE/XRT/SCIA/latest_month.html};  \cite{Takeda2016}) taken closest in time to the DAXSS observations (Fig.\ref{fig:hinode-full-images} and Appendix ~\ref{fig:hinode-full-image-faint} which enhances fainter features in the Flaring Sun observation). These are the images from April 25, 2022 at 02:06 and June 29, 2022 at 18:03 (pink vertical bands in Fig.\ref{fig:goes-lcs}). The April 25 image was taken with the Be\_thin filter and does not require any corrections for light-leaks and is therefore appropriate for use in this work as-is, whereas for the June 29 observation which was taken with the Al\_mesh filter, we needed only to apply a light leak correction. We did this using the XRTpy function `remove\_lightleak` in Python (XRTpy (v0.5.0) for {\it Hinode} X-Ray Telescope data analysis, available at \url{https://xrtpy.readthedocs.io/en/latest/};  \cite{Velasquez2024}). 

\begin{figure} 
\centering
\includegraphics[width=\columnwidth]{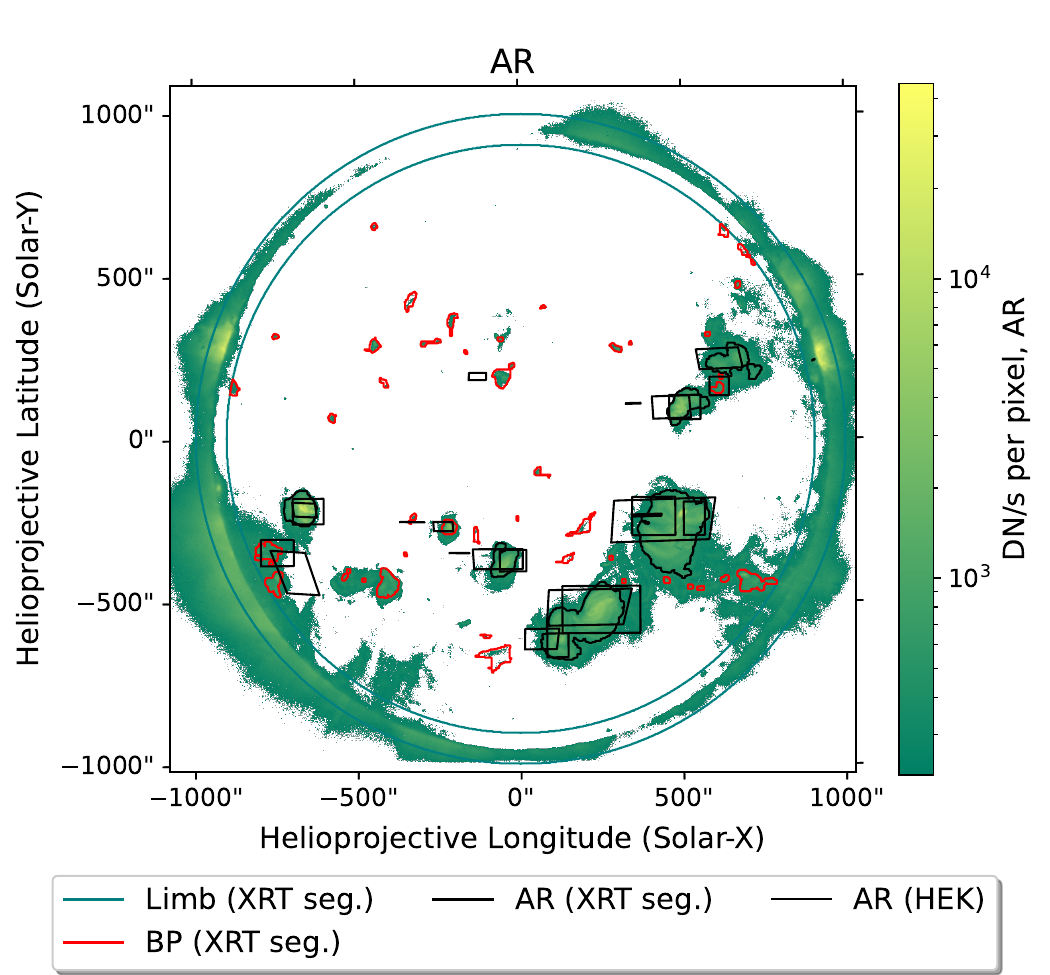}
\caption{{\it Hinode}/XRT image during Quiescent Sun observation with highlighted pixels showing the AR component (green) (Sect.\ref{sec:ff-validation-hinode}) identified from spectral fitting in Sect.\ref{subsubsec:spec-fitting-qui}. The solar limb boundaries, defined by the green circles, are taken from the {\it Hinode} segmentation maps. The inner boundary defines the reference area used in the filling factor calculations.
The black and red contours correspond to active regions and bright points, respectively, as identified in the {\it Hinode}/XRT segmentation database \citep{Adithya2021}. 
The active regions simultaneously detected by various solar instruments and registered in the public Heliophysics Event Knowledgebase (HEK; \citealt{Hulbert2012}) are shown as black bounding boxes. }.
\label{fig:hinode-obs1-ar}
\end{figure}

Since the different types of coronal regions are defined based on  {\it Yohkoh} pixel intensities ($I_{\rm BKC} < I_{\rm AR} < I_{\rm CO} < I_{\rm FL}$), we can identify corresponding regions in {\it Hinode} images by progressively masking pixels based on the filling factors derived from the spectra. We mask the brightest pixels corresponding to each region type's filling factor, starting with the hottest and brightest component and working downward in temperature. For example, the best-fit model filling factors of the flare spectrum, we designate the brightest 0.062\% of pixels as belonging to flares. Then we exclude these and designate the brightest 4.07\,\% of the remaining pixels as belonging to cores, and further excluding these, we assign the next brightest 47.51\% of pixels to active regions. 
Figures ~\ref{fig:hinode-obs1-ar} and~\ref{fig:hinode-obs2-regions} visualize the areas on the Sun identified this way for the Quiescent and the Flaring Sun observation, respectively. Appendix ~\ref{fig:hinode-obs2-regions-zoom} shows a zoom-in to the flaring region. The different regions detected with the SaXS method  are marked in  Figs.~\ref{fig:hinode-obs1-ar} and \ref{fig:hinode-obs2-regions} with  different  colorbars.  The rest of the solar corona (white in Figs.~\ref{fig:hinode-obs1-ar} and \ref{fig:hinode-obs2-regions}) is covered by the fainter types of structures (e.g. BKC for the case of the flare spectrum) which do not contribute significantly to the spectrum.

To validate our image decomposition, we compared the regions identified through the SaXS-derived filling factors with independent solar feature catalogs. Specifically, we used bounding boxes from the Heliophysics Event Knowledgebase (HEK) \citep{Hulbert2012} for active regions and flares (which HEK labels as AR and FL, respectively)---shown as black and blue boxes in Fig. ~\ref{fig:hinode-obs1-ar} and ~\ref{fig:hinode-obs2-regions}---and segmentation maps from the {\it Hinode}/XRT segmentation database \footnote{\url{https://hinode.isee.nagoya-u.ac.jp/xrt_seg/Database/XRT_database/}} \citep{Adithya2021} for active regions and bright points (labeled as AR and BP in the database), shown as colored contours in Fig. ~\ref{fig:hinode-obs1-ar} and ~\ref{fig:hinode-obs2-regions}. The {\it Hinode}/XRT segmentation database also provides inner and outer solar limb boundaries. As the features stored in both databases are restricted to within the inner solar limb, we adopt the inner boundary to define the reference area for the calculation of filling factors from their region definitions. While these catalogs use the same acronyms as our SaXS region types, their definitions differ: catalog features are identified morphologically in images, while our spectral components are defined by their characteristic thermal structure.

The filling factors we derived from the information given in the databases are summarized together with those from our SaXS spectral analysis in Table ~\ref{tab:ff-comparison}. For the Flaring Sun observation (2022 April 25), the HEK bounding boxes yield an AR filling factor almost three times higher than that obtained from the XRT database segmentation. This discrepancy primarily arises from the different instruments and region-identification methods employed in the two datasets. A smaller contribution to the difference may come from a temporal mismatch: the XRT segmentation by \citet{Adithya2021} corresponds to 05:48 UT, about three hours after the synoptic {\it Hinode} composite at 02:06 UT, by which time the flare had already subsided (see the GOES light curve in Fig.~\ref{fig:goes-lcs}). This time offset explains the slight spatial displacement between the XRT segmentation contours and the regions visible in the {\it Hinode} image.

For the Quiescent Sun observation (2022 June 29), the HEK bounding boxes again give higher AR filling factors than the {\it Hinode} segmentation maps. In this case, temporal effects are negligible, since the Sun was in a quiet phase and no major activity changes occurred between the observations. The discrepancy instead reflects the use of different instruments and different classification techniques in the two datasets. Absolute agreement between the approaches is therefore not expected, owing to these intrinsic methodological and instrumental differences.

Comparing the information from the HEK bounding boxes and XRT image segmentations to our image decomposition based on the SaXS spectral modelling  shows that the same dominant coronal structures (active regions and flaring sites) are consistently identified across all approaches. However, our SaXS spectral filling factors are systematically significantly higher than those derived from the data bases. This is because HEK data and XRT segmentations are restricted to features within the solar limb. In contrast, our SaXS-derived maps include emission extending up to one solar scale height above the limb, to better capture the total coronal output that would be observed from an unresolved stellar perspective. The additional emission we capture with SaXS analysis on full-disk spectra can be identified in Figs.~\ref{fig:hinode-obs1-ar} and \ref{fig:hinode-obs2-regions} as the colored areas outside the inner circle that denotes the solar limb. It is seen easily that a high fraction of AR emission comes from beyond the limb. We discuss these results further in Sect.~\ref{subsect:discussion_2}.

\begin{figure*} 
\centering
\parbox{\textwidth}{
\parbox{0.95\textwidth}{
\includegraphics[width=0.95\textwidth]{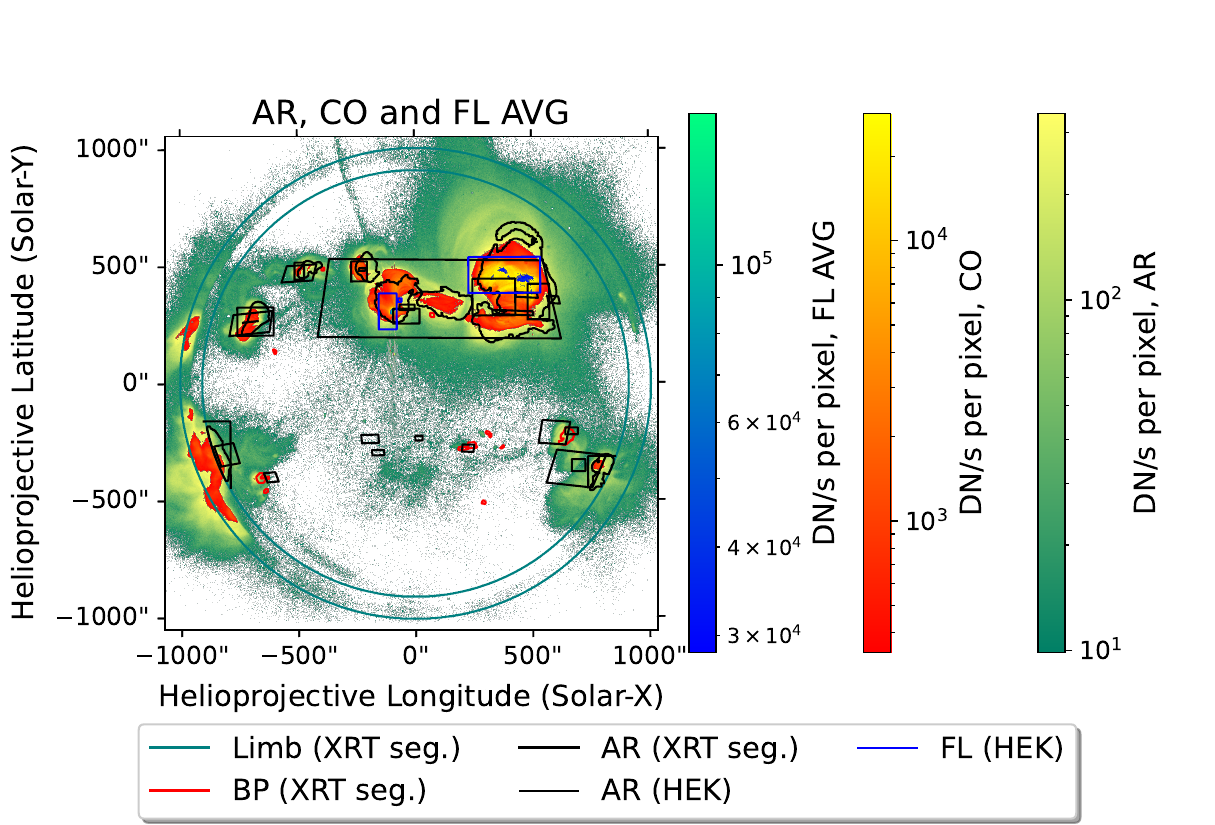}
}
}
\caption{
{\it Hinode}/XRT image taken during the Flaring Sun observation, showing pixels color-coded by coronal regions (Sect.~\ref{sec:ff-validation-hinode}) whose filling factors were determined from spectral fitting in Sect.~\ref{subsubsec:spec-fitting-flare}: AR (green), CO (orange), and FL (blue).
The solar limb boundaries (Limb (XRT seg.)), defined by the green circles, are taken from the {\it Hinode} segmentation maps. The inner boundary defines the reference area used in the filling factor calculations. 
The black and red contours correspond to active regions (AR (XRT seg.)) and bright points (BP (XRT seg.)), respectively, as identified in the {\it Hinode}/XRT segmentation database \citep{Adithya2021}. 
The flares (FL (HEK)) and active regions (AR (HEK)) simultaneously detected by various solar instruments and registered in the public Heliophysics Event Knowledgebase (HEK; \citealt{Hulbert2012}) are shown as black and blue bounding boxes, respectively.
}.
\label{fig:hinode-obs2-regions}
\end{figure*}

\begin{table*}
  \begin{center}
    \caption{Comparison of coronal-region filling factors (\%) derived from SaXS spectral analysis and from image-based catalogs.}
    \label{tab:ff-comparison}
    \begin{tabular}{l|l|l|l|l|l}
      \textbf{Observation Date} & \textbf{Region Type} & \textbf{SaXS (this work)} & \textbf{HEK}$^{a}$ & \textbf{\textit{Hinode}/XRT}$^{b}$ & \textbf{Notes} \\
      \hline
      \multirow{3}{*}{2022 Apr 25 (Flaring Sun)} 
      & AR & $47.51 \pm 13.97$ & 19.91 & 7.66 (7.93\,AR+BP) & Flaring Sun DAXSS spectrum \\
      & CO          & $4.07 \pm 0.14$ & --- & --- & Not segmented in image databases \\
      & FL MED         & $0.062 \pm 0.00072$ & 2.25 & --- & Only from HEK bounding boxes \\
      \hline
      \multirow{2}{*}{2022 Jun 29 (Quiescent Sun)} 
      & AR & $21.83 \pm 0.40$ & 8.06 & 4.02 & Quiescent Sun DAXSS spectrum \\
      & BKC & $94.12 \pm 65.85$ & --- & --- & Not segmented in image databases \\
      \hline
    \end{tabular}
  \end{center}
  \vspace{-0.3cm}
  \begin{flushleft}
    \small
    \textbf{Notes.} Filling factors (FF) represent the fractional area of the solar disk within the solar limb (as defined by the {\it Hinode}/XRT segmentation maps) occupied by each coronal region.  
    $^{a}$HEK: bounding boxes of active regions and flares from multiple solar missions \citep{Hulbert2012}.  
    $^{b}$\textit{Hinode}/XRT: segmentation maps of active regions and bright points from \citet{Adithya2021}.  
  \end{flushleft}
\end{table*}

\section{Discussion}
\label{sec:discussion}

\subsection{The new SaXS method: a physical description of spectra and dynamics in stellar coronae}
\label{subsect:discussion_1}

We have developed a fundamentally  new approach  to the Sun-as-an-Xray-star prescription for comparing stellar X-ray observations to solar data. Our new implementation consists in creating X-ray spectral models for the individual types of solar coronal regions. Specifically we define {\sc XSPEC} models for BKC, AR, CO and FL-AVG, by converting their {\it Yohkoh} EMDs each into  a multi-temperature {\sc apec} model in which each temperature-component is appropriately weighted. These {\sc XSPEC} models can replace previous uninformed spectral fitting yielding a physical description of the emitting plasma in the corona of magnetically active stars. We have demonstrated this here on the example of the Sun itself. 

Previously, quiescent and flaring solar spectra from DAXSS have been analyzed using one- or two-temperature isothermal models with fixed or variable abundances \citep[e.g.,][]{Schwab2020,Schwab2023}. These models provide a reasonable approximation of the integrated emission from different coronal regions that cannot be directly resolved on other stars, and they are standard practice in stellar X-ray spectroscopy. \citet{Schwab2020} found that the hotter temperature component in their two-temperature fits was associated with active regions visible in contemporaneous solar images. Their quiescent Sun fits yielded $\chi^2_{\rm red} \sim 7.7$ with a $1T$ model and $\sim4.5$ with $2T$ components, improving further to $\sim3.9$ when allowing elemental abundances to vary in the fit. 

These results are comparable to $\chi^2_{\rm red}$ values of $5.33$ and $2.26$ that we obtain for our SaXS model fits on the Quiescent and Flaring Sun spectra, respectively, despite keeping abundances fixed. This shows that the current version of SaXS spectral models achieve fits of similar quality to standard multi-temperature models while additionally providing a physical link between spectral components and specific types of coronal structures.
Specifically, it enables direct retrieval of coronal filling factors. Further improvements of the SaXS models are foreseen in the future as we discuss in Sect.~\ref{subsect:discussion_3}

\subsection{Consistency check of the new SaXS approach on solar data}
\label{subsect:discussion_2}

In the first application of the new SaXS method we have tested the technique on the Sun itself. We have selected two DAXSS spectra, one representing the Quiescent and one the Flaring Sun.
We found that the Quiescent Sun spectrum was best fit with the BKC and AR models (Fig.~\ref{fig:bkc-ar-dist-reduced-spectra}, bottom panel) and the Flaring Sun spectrum was best fit with AR, CO and FL-AVG regions (Fig.~\ref{fig:fl-obs-fit}, bottom panel). The filling factors derived from the SaXS spectral fitting were then visualized on {\it Hinode} images as described in Sect.~\ref{sec:ff-validation-hinode}. 
To validate our results, we compared the regions identified through our SaXS spectral analysis on the {\it Hinode} images with image segmentations carried out by others.

Overall, we found good spatial correspondence between the regions detected by our method and those listed in the HEK and the {\it Hinode}/XRT databases, although the filling factors do not match exactly. There are several reasons for this. First, both HEK and {\it Hinode} segmentations are strictly limited to features within the solar limb, whereas we also include emission from within one solar scale height above the limb. The location of the solar limb itself was adopted from the {\it Hinode} segmentation maps and used as the reference area in our filling-factor calculations. This choice reflects the stellar-analogy goal of our study—i.e., to account for all coronal emission that would be observed if the Sun were seen as an unresolved star. Second, HEK regions are identified using a variety of instruments and wavelengths, each with its own detection and validation procedures, while the {\it Hinode} maps are based solely on broadband X-ray data. Our approach, in contrast, relies primarily on intensity thresholds applied uniformly to broad-band images. Finally, some categories used in our analysis (e.g., CO regions) are 
not 
explicitly listed in current solar feature databases, further limiting one-to-one correspondence. Despite these differences, the same major coronal structures are consistently detected by all methods, indicating that the filling factors inferred from SaXS are physically meaningful.

\subsection{Open issues}
\label{subsect:discussion_3}

Despite the compelling qualitative results, several caveats and issues to be studied in future work have been identified and we discuss them briefly in the following.

First, we are currently restricted to applying our spectral models with a low-energy cutoff of 0.7\,keV because DAXSS is not well calibrated below this threshold \citep{Schwab2020,Schwab2023}. This means that the energy range of 0.2–0.7\,keV which is an important part of the spectra provided by stellar X-ray observatories, e.g. {\it XMM-Newton} and eROSITA,   remains untested. 
This soft energy regime is essential for constraining the spectral contribution of low-temperature coronal structures like BKC and AR. In fact, in our tests on the DAXSS spectra the filling factor of the components contributing the highest emission to the spectra  can be constrained well (see Sect.~\ref{sec:spec-fitting}), but those contributing little to the  emission can be constrained only poorly, even if they occupy the largest projected area on the corona. 
For example, in the Quiescent Sun spectrum the fit improves only slightly when the BKC component is added, whereas in the Flaring Sun spectrum the inclusion of BKC worsens the fit. This does not imply that these regions are absent from the solar disk at the time of observation; rather, their emission measure per unit area is too low for their contribution to be detectable in the integrated spectrum with its limitation to $E > 0.7$\,keV. Consequently, our method currently provides reasonable constraints for the higher temperature and more emissive coronal structures, but is less sensitive to low-emission and low-temperature regions.

Alternatively, the remaining residuals we see could reflect physical effects not captured by the current equilibrium models, such as departures from ionization equilibrium or an additional thermal component in the solar corona that is not represented by the current set of SaXS models. Future fitting tests incorporating non-equilibrium ionization or adding a simple isothermal component could help determine which plasma characteristics are responsible for the mismatch.

In order to obtain a reasonably good fit to the DAXSS spectra, meaning that systematic trends in residuals over a large energy range are avoided, we had to modify ad hoc the original {\it Yohkoh} EMDs. The restriction of the EMDs,  illustrated in Appendix.~\ref{fig:em-dist-reduced} and described in Sect.~\ref{subsect:refining-emd}, is purely based on the observation that both the Quiescent Sun and the Flaring Sun spectra have steeper high-energy drop offs than any of the spectral models that were created using the entire original EMDs, as illustrated in Fig.~\ref{fig:demo-all-em-bins-bad-fit}. 
A possible physical reason for the need of such an ad hoc adjustment may lie in the fact that the {\it Yohkoh} EMDs are based on only a handful of {\it Yohkoh} observations. Our EMDs are, therefore, not necessarily representative for the typical case of each type of coronal structure. 
The discrepancy also likely arises because the {\it Yohkoh} EMDs derived by \citealt{Orlando2001}, \citealt{Orlando2004} and \citealt{Reale2001} represent averages from individual EMDs, whereas the DAXSS data reflect a specific configuration of regions that may differ from this average. \cite{Orlando2004} have shown that the high-temperature tail of AR and CO strongly depends on their activity level: newly emerged AR  and CO  exhibit a pronounced high-temperature tail, while more evolved or less AR and CO display much weaker high-temperature  emission. The fact that to fit the DAXSS spectra it is necessary to suppress the high-temperature tail of the EMDs suggests that the structures on the solar corona during the time of DAXSS observation were evolved and with low levels of activity. A larger set of $EMD_{\rm reg}$ describing different states of the individual region types should be derived for more precise results from the SaXS spectral fitting. 

With the empirically  "restricted EMDs" the overall slope of both the Quiescent and Flaring Sun can be described with suitable combinations of BKC, AR, CO and FL-AVG. However, the $\chi^2_{\rm red}$ remains high and systematic residuals are present that can be associated to individual emission lines (see \citet{ Woods2023,Telikicherla2024}). This means that in future studies, during the generation of the spectral models, the individual abundances of various elements will have to be modified. Clearly, this will add additional complexity to the fitting process, and it requires a dedicated investigation.

\subsection{Outlook at application of the new SaXS approach to stellar X-ray spectra}

The ultimate goal of our work is to apply the new SaXS implementation to stellar X-ray spectra, and derive coronal filling factors for unresolved stellar surfaces. The SaXS method inherently assumes that the EMDs of coronal regions we observe on other magnetically active stars are the same as those on the Sun. There are good reasons for this assumption,  since stellar EMDs have been shown to occupy the same temperature ranges and have similar shapes  to what we observe on the Sun \citep{Jordan2012,Peres2004}.
In Joseph et al., (in prep.) we present a first application of the new SaXS implementation to the flare star AD\,Leo. There we explain also how the way of deriving filling factors (as described in Sect.~\ref{sec:model-creation} for the solar case) needs to be adjusted for stars to account for the different densities in their corona.

\section{Conclusions}
\label{sec:conclusions}
We have developed and tested a new implementation of the SaXS method that converts observed solar coronal region EMDs into spectral models to be used within {\sc XSPEC}. They can, therefore, be used  to fit stellar and solar X-ray spectra providing as main output the filling factors of the different types of magnetic structures in the corona. In this work we focused on validating the new SaXS approach on the Sun. Applied to DAXSS observations of the Quiescent and Flaring Sun, the SaXS spectral models qualitatively reproduce the spectra reasonably well. Their predicted filling factors were validated against {\it Hinode} X-ray images.
Our results show that the filling factors of dominant emitting regions in the solar corona (e.g., ARs during quiescence and AR+CO+FL-AVG during flares) can be constrained robustly, while weaker-emission regions remain less well constrained. 

This work presents the first implementation of SaXS with separate spectral models for each type of coronal region, tested directly against solar spectra. 
The SaXS spectral models thus provide a practical bridge between solar observations and unresolved stellar coronae, offering an alternative to traditional few-temperature {\sc apec} fitting. While empirical, the approach is grounded in the similarity of stellar and solar EMDs, and allows direct retrieval of coronal region filling factors for spatially unresolved surfaces of active stars. Thus, it opens a path toward quantifying stellar activity in terms of the relative abundances of distinct coronal structures, rather than relying on purely phenomenological global plasma temperatures. 

We validate our spectrally-derived filling factors against the HEK and {\it Hinode} databases of independently classified solar regions. These datasets provide a valuable cross-check, although their spatial definitions and temporal coverage differ substantially.  
The HEK entries are derived from multiple instruments and wavelengths, while the {\it Hinode}/XRT segmentations are based solely on broadband X-ray data. The inclusion of the emission from the limb together with differences in the definition of the  boundaries between the different region types explains the larger filling factor we obtain with SaXS compared to the results in the two databases.  

The limitations of the present work include the low-energy  response cutoff of DAXSS at 0.7\,keV and the reliance on a small set of {\it Yohkoh}-derived EMDs which do not systematically sample all solar magnetic region types.
Nevertheless, the method provides a framework that can be refined and tested with larger solar samples (making use of contemporaneous {\it Hinode}, DAXSS, and GOES data). Already, in its present version, its application to stars provides interesting insight (Joseph et al., in prep.).

In summary, this work introduces spectral SaXS as a viable tool for studying magnetically active stars, enabling direct inference of coronal region filling factors from X-ray spectra and providing a new approach to characterizing stellar activity.

\begin{acknowledgements}

WMJ acknowledges support by the Bundesministerium für Wirtschaft und Energie through the Deutsches Zentrum für Luft- und Raumfahrt e.V. (DLR) under grant numbers FKZ 50 OR 2208. DAXSS (INSPIRESat-1) solar soft X-ray spectral irradiance data is provided by LASP/University of Colorado (MinXSS/DAXSS team); we used version 2.1.0  of the Level 2 data product. Hinode is a Japanese mission developed and launched by ISAS/JAXA, collaborating with NAOJ as a domestic partner, NASA and STFC (UK) as international partners. Scientific operation of the Hinode mission is conducted by the Hinode science team organized at ISAS/JAXA. This team mainly consists of scientists from institutes in the partner countries. Support for the post-launch operation is provided by JAXA and NAOJ (Japan), STFC (U.K.), NASA, ESA, and NSC (Norway).

\end{acknowledgements}

\bibliographystyle{bibtex/aa.bst} 
\bibliography{biblio}
\FloatBarrier

\begin{appendix}

\section{SaXS spectral models and their components}

Figure~\ref{fig:model-apec-comps} illustrates the SaXS spectral models implemented in {\ XSPEC}. They were constructed from the emission measure distributions (EMDs) of the different coronal region types defined in Sect.~\ref{sec:model-creation}.  
Each panel shows the composite model spectrum (colored line) together with the individual {\sc apec} components (grey curves) that represent the isothermal plasma contributions corresponding to the temperature bins of the respective EMD.  
The colors denote the coronal region type: red for the background corona (BKC), purple for active regions (AR), yellow for cores (CO), and blue for the averaged flare distribution (FL-AVG).  
These models form the basis of the SaXS fitting procedure described in Sect.~\ref{sec:solar-data}, where they are combined in varying proportions to reproduce  observed solar soft X-ray spectra.

\begin{figure*}[h]
\centering
\parbox{\textwidth}{
	\parbox{0.5\textwidth}{\includegraphics[width=0.5\textwidth]{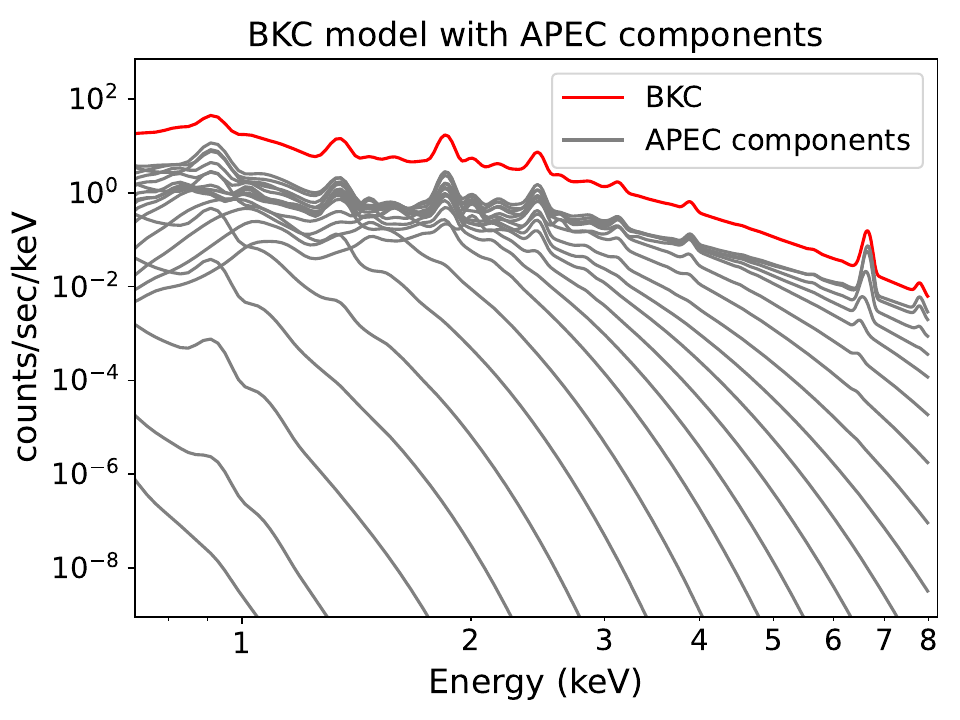}}
	\parbox{0.5\textwidth}{\includegraphics[width=0.5\textwidth]{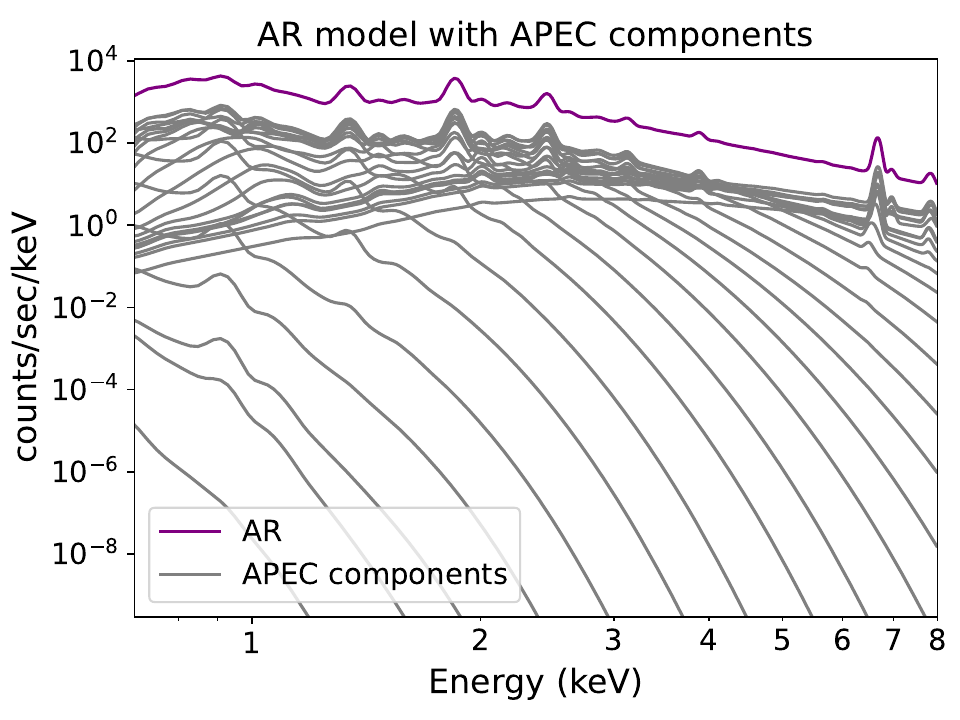}}
}
\parbox{\textwidth}{
	\parbox{0.5\textwidth}{\includegraphics[width=0.5\textwidth]{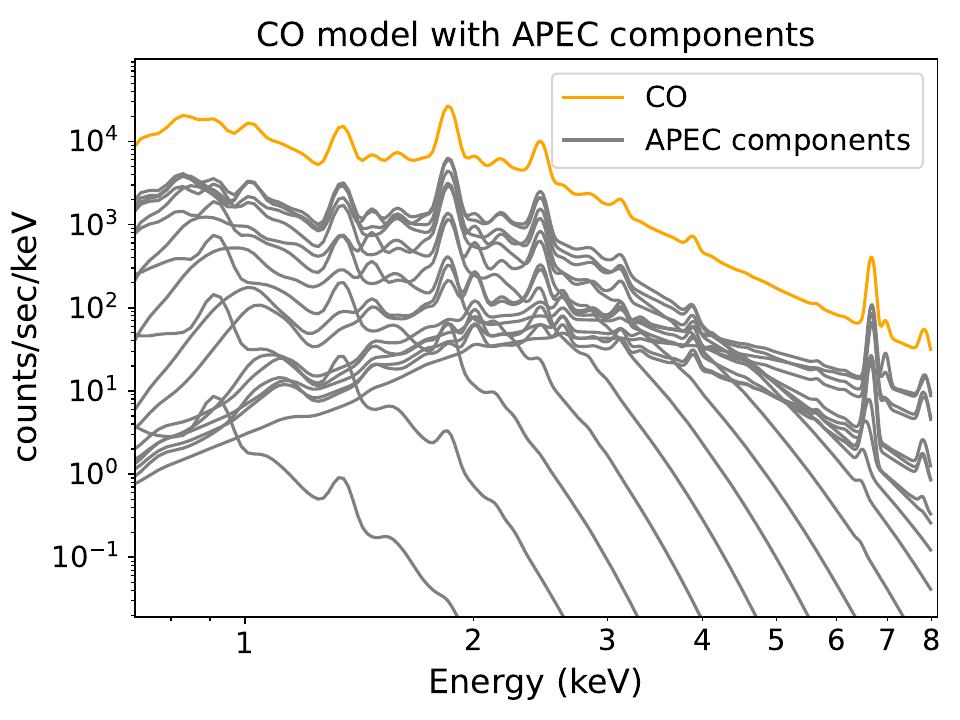}}
	\parbox{0.5\textwidth}{\includegraphics[width=0.5\textwidth]{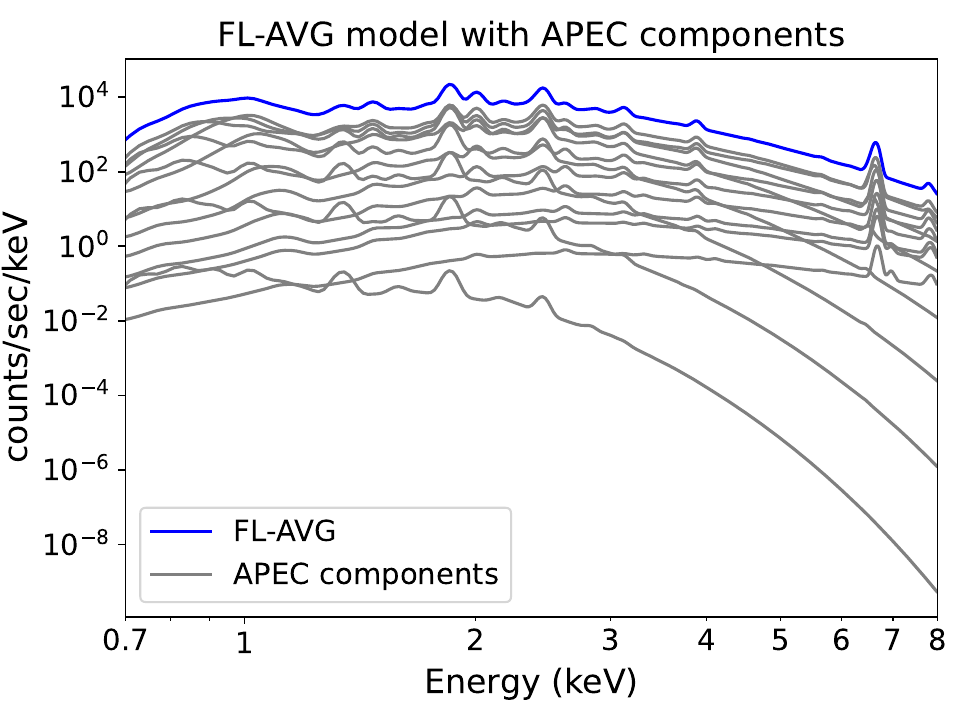}}
}
\addtocounter{figure}{0}
\caption{Composite SaXS  spectral models made with {\sc XSPEC} (colored) and their individual {\sc apec} components (grey) for the main solar coronal region types used in the SaXS method (see Sect.~\ref{sec:model-creation}).  
Colors indicate the coronal region type: red for the background corona (BKC), purple for the active regions (AR), yellow for the cores (CO), and blue for the averaged flares (FL-AVG).}
\label{fig:model-apec-comps}
\end{figure*}

\FloatBarrier

\section{Restricted vs original EMDs}

Figure~\ref{fig:em-dist-reduced} compares the original {\it Yohkoh} EMDs of the coronal region types introduced in Sect.~\ref{sect:intro} with their restricted versions used in this work (Sect.~\ref{subsect:refining-emd}).  
The restricted EMDs include only temperature bins within one order of magnitude of the peak emission measure of each distribution, effectively removing the low- and high-temperature tails that were found to overestimate the flux in the high-energy portion of the 
observed solar broad-band spectra. In Fig.~\ref{fig:em-dist-reduced} opaque lines correspond to the restricted EMDs adopted for the spectral modeling, while the transparent lines show the original distributions derived from {\it Yohkoh}/SXT data which are also presented in Fig.~\ref{fig:em-dist}.
\begin{figure}[h]
\centering
\includegraphics[width=\columnwidth]{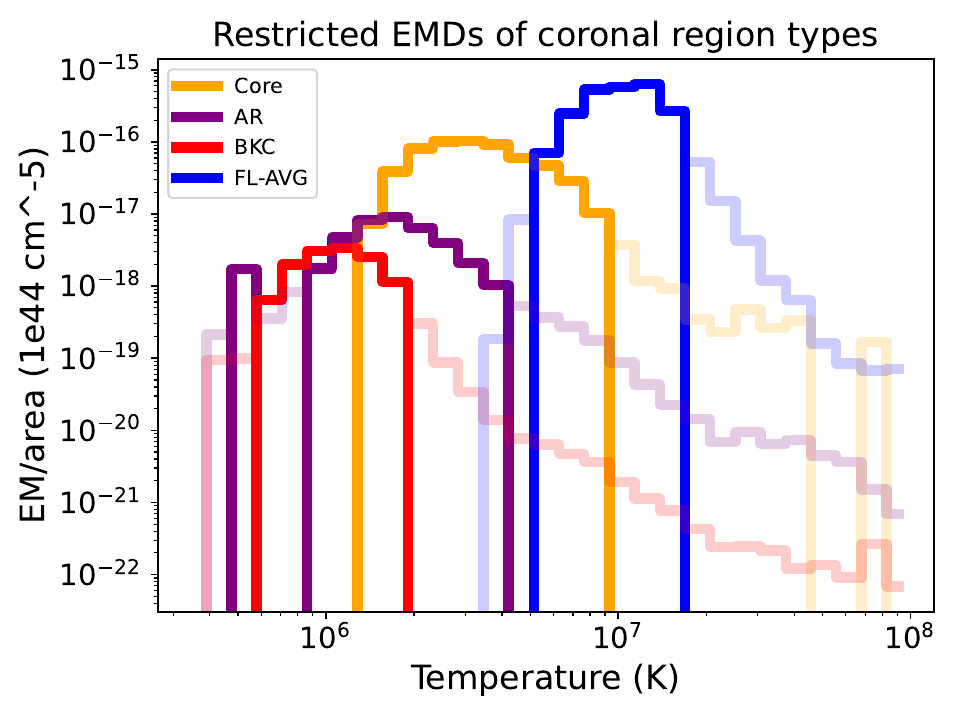}
\caption{Emission measure distributions of the BKC, AR, CO and FL-AVG regions as described in Sect.~\ref{sect:intro} the opaque lines are the restricted distributions we use in this paper (described in Sect.~\ref{subsect:refining-emd}), and the transparent lines depict the original distributions.}
\label{fig:em-dist-reduced}
\end{figure}

\FloatBarrier

\section{Comparison of SaXS spectral models from restricted vs original EMDs}

Figure~\ref{fig:spec-models-em-dist-reduced} compares the SaXS spectral models in {\sc XSPEC}obtained from the full emission measure distributions (EMDs) with those generated from the restricted EMDs introduced in Sect.~\ref{subsect:refining-emd}.  
Each panel shows the model corresponding to one coronal region type: background corona (BKC, red), active regions (AR, purple), cores (CO, yellow), and averaged flares (FL-AVG, blue).  
Solid lines represent the spectral models derived from the original full EMDs, while dashed lines show the models constructed from the reduced EMDs that exclude low- and high-temperature bins with emission measures more than one order of magnitude below the EMD peak.  
These comparisons illustrate how removing the low-signal EMD tails reduces excess high-energy flux, leading to improved agreement with the observed DAXSS spectra.

\begin{figure*}[h]
\centering
\parbox{\textwidth}{
\parbox{0.5\textwidth}{\includegraphics[width=0.5\textwidth]{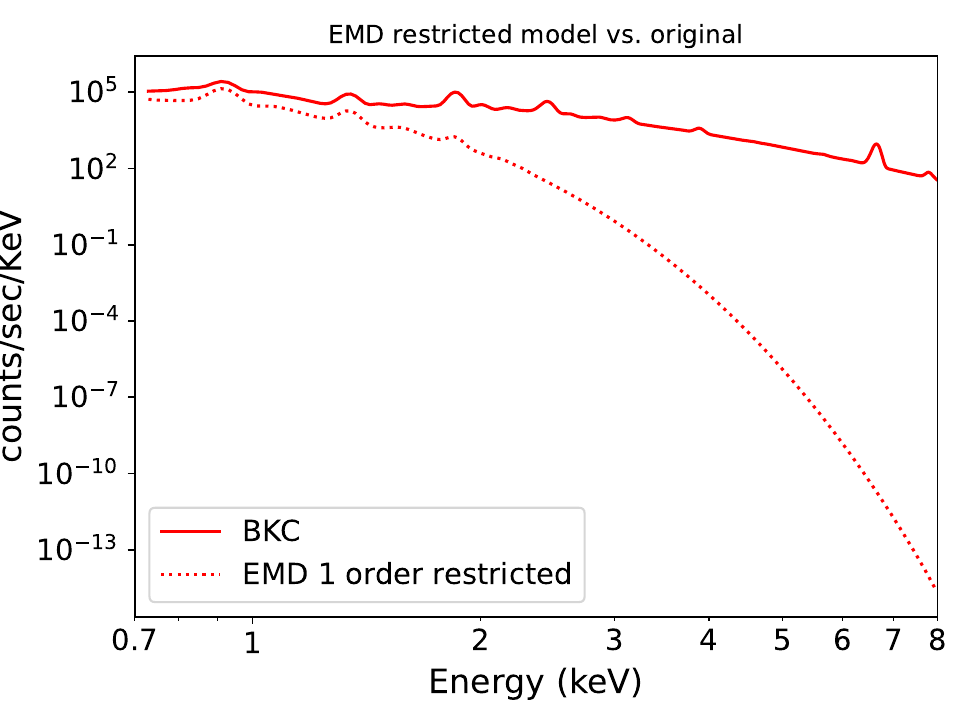}}
\parbox{0.5\textwidth}{\includegraphics[width=0.5\textwidth]{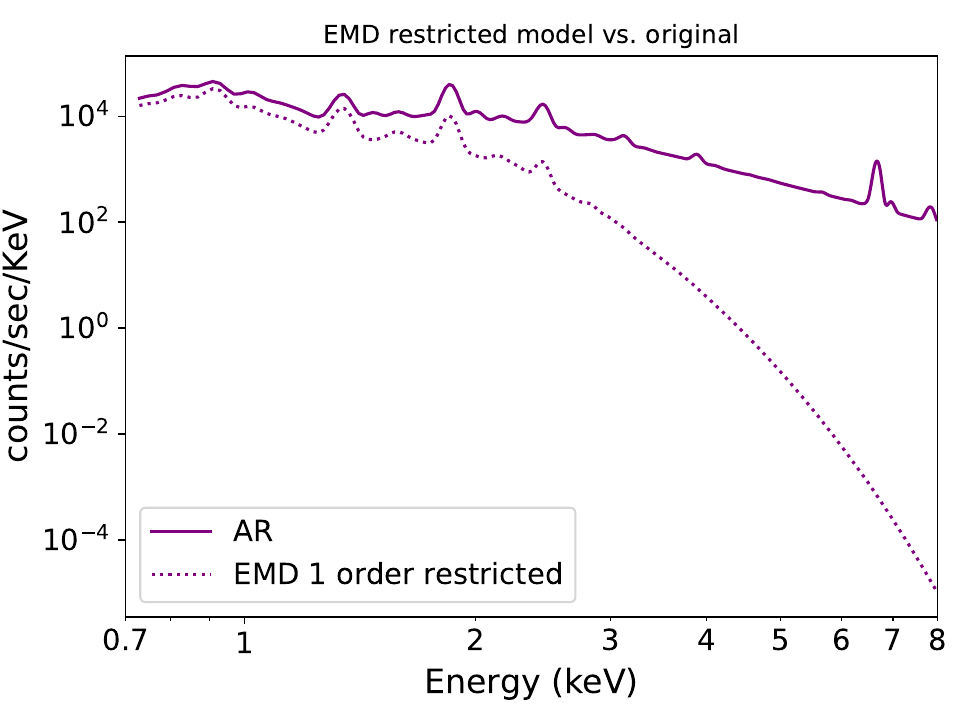}}
}
\parbox{\textwidth}{
\parbox{0.5\textwidth}{\includegraphics[width=0.5\textwidth]{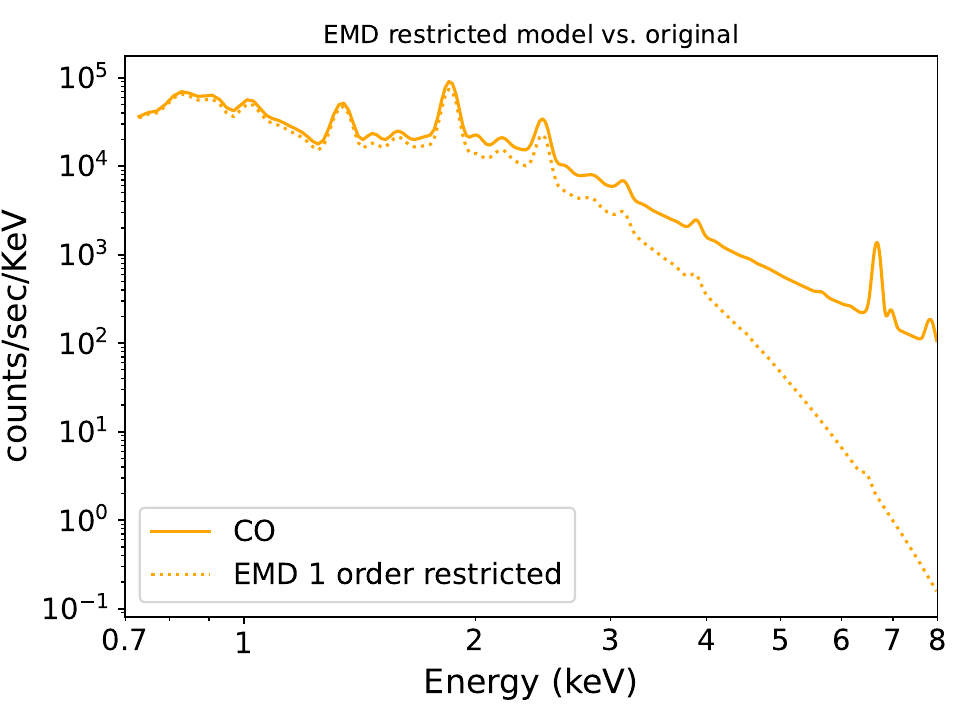}}
\parbox{0.5\textwidth}{\includegraphics[width=0.5\textwidth]{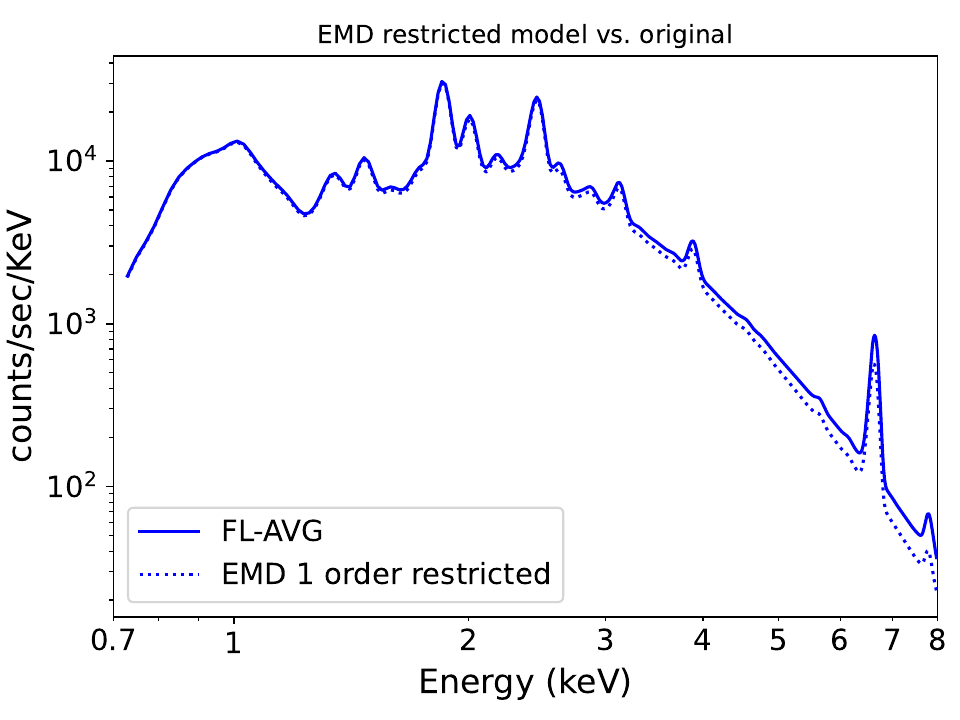}}
}
\addtocounter{figure}{0}
\caption{Comparison between composite SaXS  spectral models in {\sc XSPEC} derived from the full (solid lines) and restricted (dashed lines) emission measure distributions for the main coronal region types defined in Sect.~\ref{subsect:refining-emd}.  
Colors indicate the region type: red for the background corona (BKC), purple for active regions (AR), yellow for cores (CO), and blue for the averaged flares (FL-AVG).}
\label{fig:spec-models-em-dist-reduced}
\end{figure*}

\FloatBarrier

\section{\textit{Hinode}/XRT image of the Flaring Sun observation with faint coronal features enhanced}
\begin{figure}[h]
\centering
\includegraphics[width=\columnwidth]{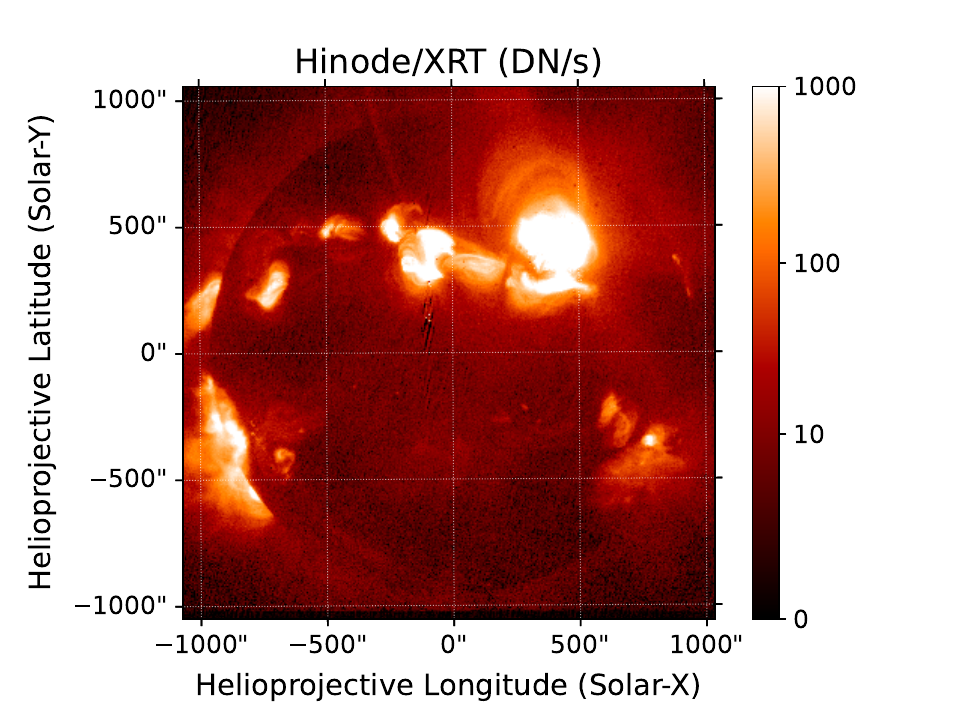}
\caption{{\it Hinode}/XRT image corresponding to the Flaring Sun DAXSS observation. The colorbar limits are adjusted to enhance visibility of coronal structures that are faint relative to the ongoing flares}.
\label{fig:hinode-full-image-faint}
\end{figure}

Figure ~\ref{fig:hinode-full-image-faint} shows the \textit{Hinode}/XRT image corresponding to the Flaring Sun observation (see Sect.~\ref{sec:ff-validation-hinode}), with the colorbar scales adjusted to show relatively faint coronal features that are otherwise masked by ongoing flares.

\FloatBarrier

\section{Close-up of flaring region on the Flaring Sun {\it Hinode} image}

\begin{figure*} 
\centering
\parbox{\textwidth}{
\parbox{0.95\textwidth}{
\includegraphics[width=\textwidth]{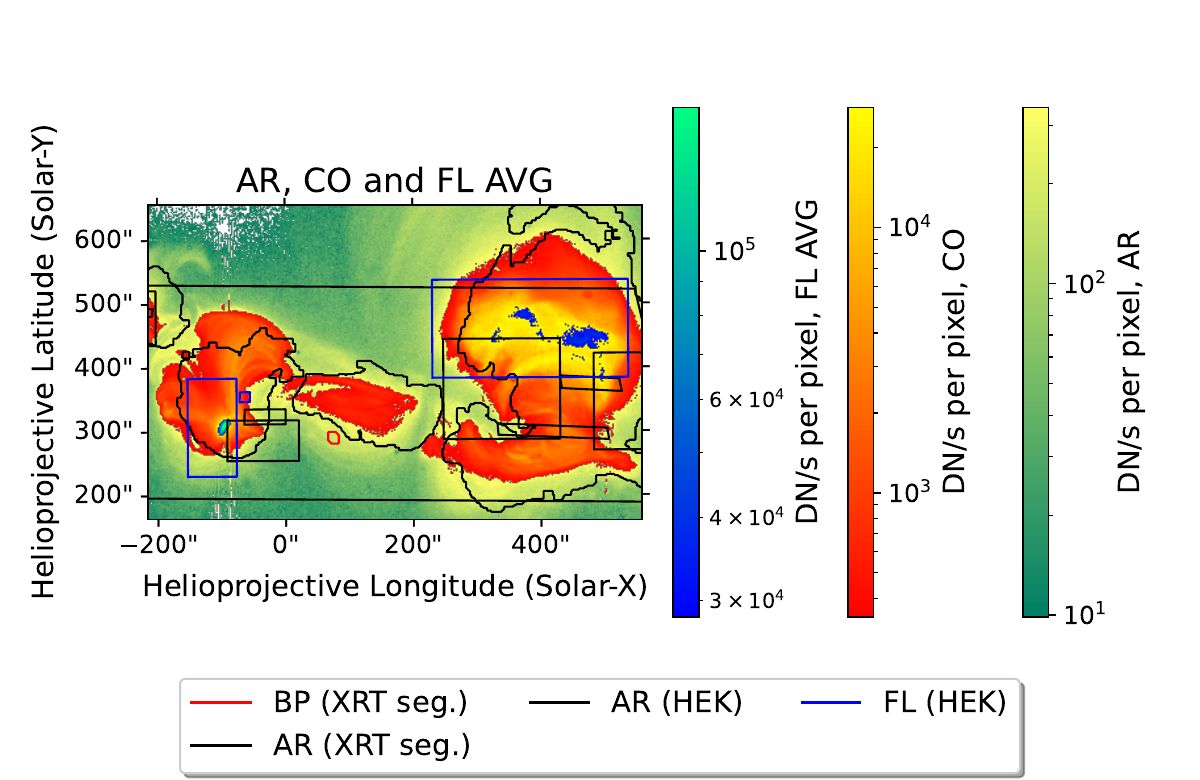}
}
}
\caption{Zoom-in to flaring regions on the {\it Hinode}/XRT image taken during the Flaring Sun observation shown in Fig.~\ref{fig:hinode-obs2-regions}, showing pixels color-coded by coronal regions (Sect.~\ref{sec:ff-validation-hinode}) whose filling factors were determined from spectral fitting in Sect.~\ref{subsubsec:spec-fitting-flare}: AR (green), CO (orange), and FL (blue). 
The black and red contours correspond to active regions (AR (XRT seg.)) and bright points (BP (XRT seg.)), respectively, as identified in the {\it Hinode}/XRT segmentation database \citep{Adithya2021}. 
The flares (FL (HEK)) and active regions (AR (HEK)) simultaneously detected by various solar instruments and registered in the public Heliophysics Event Knowledgebase (HEK; \citealt{Hulbert2012}) are shown as black and blue bounding boxes, respectively.
}.
\label{fig:hinode-obs2-regions-zoom}
\end{figure*}

Figure ~\ref{fig:hinode-obs2-regions-zoom} shows a zoom-in on the flaring regions observed by {\it Hinode}/XRT during the Flaring Sun observation (2022 April 25). Pixels are color-coded according to the coronal region types identified in Sect.~\ref{sec:ff-validation-hinode}. Their filling factors were derived from the spectral fitting in Sect.~\ref{subsubsec:spec-fitting-flare}: active regions (AR; green), cores (CO; orange), and flares (FL; blue). The contours correspond to active regions (AR (XRT seg.); black) 
and bright points (BP (XRT seg.); red), from the {\it Hinode}/XRT segmentation database \citep{Adithya2021}. Bounding boxes marking flares (FL (HEK); black) and active regions (AR (HEK); blue) detected by multiple solar instruments and registered in the Heliophysics Event Knowledgebase (HEK; \citealt{Hulbert2012}) are also shown. The figure illustrates the correspondence between spectrally derived coronal components and regions identified in imaging-based catalogs.

\end{appendix}
\end{document}